\documentclass[12pt]{article}
\usepackage{latexsym}
\usepackage{amssymb}
\usepackage{graphicx}

\def\be{\begin{equation}}
\def\ee{\end{equation}}
\def\bear{\begin{eqnarray}}
\def\eear{\end{eqnarray}}
\def\nn{\nonumber}

\newcommand\bra[1]{{\langle {#1}|}}
\newcommand\ket[1]{{|{#1}\rangle}}

\def\a{\alpha}
\def\b{\beta}
\def\g{\gamma}
\def\d{\delta}

\def\r{\rho}
\def\t{\tau}

\def\s{\sigma}
\def\th{\theta}

 \def\o{{\rm ord}}
 
 \def\L{{\Lambda}}
 \def\CN{{\cal N}}

 \def\r{\rightarrow}

 \def\IZ{\relax\ifmmode\mathchoice
 {\hbox{\cmss Z\kern-.4em Z}}{\hbox{\cmss Z\kern-.4em Z}}
 {\lower.9pt\hbox{\cmsss Z\kern-.4em Z}}
 {\lower1.2pt\hbox{\cmsss Z\kern-.4em Z}}\else{\cmss Z\kern-.4em Z}\fi}
 \def\IB{\relax{\rm I\kern-.18em B}}
 \def\IC{{\relax\hbox{$\inbar\kern-.3em{\rm C}$}}}
 \def\Ic{{\relax\hbox{$\inbar\kern-.22em{\rm c}$}}}
 \def\ID{\relax{\rm I\kern-.18em D}}
 \def\IE{\relax{\rm I\kern-.18em E}}
 \def\IF{\relax{\rm I\kern-.18em F}}
 \def\IG{\relax\hbox{$\inbar\kern-.3em{\rm G}$}}
 \def\IGa{\relax\hbox{${\rm I}\kern-.18em\Gamma$}}
 \def\IH{\relax{\rm I\kern-.18em H}}
 \def\II{\relax{\rm I\kern-.18em I}}
 \def\IK{\relax{\rm I\kern-.18em K}}
 \def\IP{\relax{\rm I\kern-.18em P}}

\def\Tr{{\rm Tr}}
 \font\cmss=cmss10 \font\cmsss=cmss10 at 7pt
 \def\IR{\relax{\rm I\kern-.18em R}}

\def\dd{\mbox{d}}

\def\O{\Omega}
\def\o{\omega}
\def\bra{\langle}
\def\ket{\rangle}
\def\a{\alpha}
\def\b{\beta}
\def\d{\delta}
\def\D{\Delta}

\def\g{\gamma}
\def\G{\Gamma}
\def\e{\epsilon}
\def\ve{\varepsilon}

\def\f{\phi}

\def\vf{\varphi}

\def\l{\lambda}
\def\L{\Lambda}
\def\m{\mu}
\def\n{\nu}
\def\s{\sigma}
\def\S{\Sigma}
\def\o{\omega}

\def\r{\rho}
\def\t{\tau}
\def\th{\theta}

\def\pa{\partial}

\newcommand{\ti}[1]{\tilde{#1}}

\newcommand{\sm}[1]{\mbox{\scriptsize #1}}

\renewcommand{\@}[1]{\sqrt{#1}}
\renewcommand{\le}[1]{\label{#1}\end{eqnarray}}
\newcommand{\bea}{\begin{eqnarray}}
\newcommand{\eea}{\end{eqnarray}}

\newcommand{\eq}[1]{(\ref{#1})}
\def\nn{\nonumber\\}

\def\ffract#1#2{\raise .35 em\hbox{$\scriptstyle#1$}\kern-.25em/
\kern-.2em\lower .22 em \hbox{$\scriptstyle#2$}}

\def\na{\nabla}
\def\half{{1\over2}\,}
\newdimen\tableauside\tableauside=1.0ex
\newdimen\tableaurule\tableaurule=0.4pt
\newdimen\tableaustep
\def\phantomhrule#1{\hbox{\vbox to0pt{\hrule height\tableaurule width#1\vss}}}
\def\phantomvrule#1{\vbox{\hbox to0pt{\vrule width\tableaurule height#1\hss}}}
\def\sqr{\vbox{%
  \phantomhrule\tableaustep
  \hbox{\phantomvrule\tableaustep\kern\tableaustep\phantomvrule\tableaustep}%
  \hbox{\vbox{\phantomhrule\tableauside}\kern-\tableaurule}}}
\def\squares#1{\hbox{\count0=#1\noindent\loop\sqr
  \advance\count0 by-1 \ifnum\count0>0\repeat}}
\def\tableau#1{\vcenter{\offinterlineskip
  \tableaustep=\tableauside\advance\tableaustep by-\tableaurule
  \kern\normallineskip\hbox
    {\kern\normallineskip\vbox
      {\gettableau#1 0 }%
     \kern\normallineskip\kern\tableaurule}%
  \kern\normallineskip\kern\tableaurule}}
\def\gettableau#1 {\ifnum#1=0\let\next=\null\else
  \squares{#1}\let\next=\gettableau\fi\next}

\makeatletter
\@addtoreset{equation}{section}
\makeatother

\begin{document}

\begin{flushright}
KCL-MTH-07-02\\
UPR-1163-T\\
\end{flushright}
\vskip0.1truecm

\begin{center}
\vskip 1truecm {\Large \textbf{Electric-magnetic Duality \\
\vskip .3truecm
and Deformations of Three-Dimensional CFT's}}
\vskip 2truecm

{\large Sebastian de Haro${}^\star$ and Peng Gao${}^\dagger$}\\

\vskip .4truecm
\vskip 4truemm ${}^\star${\it Department of Mathematics\\
King's College, London WC2R 2LS, UK}\\
\tt{sebastian.deharo@kcl.ac.uk}\\
\vskip .3truecm
$^\dagger${\it Department of Physics and Astronomy\\
University of Pennsylvania\\
Philadelphia, PA 19104-6396, USA}\\
{\tt gaopeng@physics.upenn.edu}

\end{center}
\vskip 1truecm

\begin{center}

\textbf{\large \bf Abstract}

\end{center}

$SL(2,{\mathbb{Z}})$ duality transformations in asymptotically AdS$_4\times S^7$ act non-trivially on the
three-dimensional SCFT of coincident M2-branes on the boundary. We show how $S$-duality acts away from the
IR fixed point. We develop a systematic method to holographically obtain the deformations 
of the boundary CFT and show how electric-magnetic duality relates different deformations.
We analyze in detail marginal deformations and deformations by dimension 4 operators. In 
the case of massive deformations, the RG flow relates $S$-dual CFT's. Correlation functions in the CFT are 
computed by varying magnetic bulk sources,
whereas correlation functions in the dual CFT are computed by electric bulk sources. Under 
massive deformations, the boundary 
effective action is generically minimized by massive self-dual configurations of the 
$U(1)$ gauge field. We show that a self-dual choice of boundary conditions exists, and it 
corresponds to the self-dual topologically massive gauge theory in 2+1 dimensions. Thus, self-duality in 
three dimensions can be understood as a consequence of electric-magnetic invariance in the 
bulk of AdS$_4$.

\newpage

\tableofcontents


\section{Introduction and summary of the results}

Electric-magnetic duality has played an important role in understanding non-perturbative 
aspects of supersymmetric gauge theories and string theory. It was instrumental in the description of 
confinement by monopole condensation in ${\cal N}=2$ SYM theory by Seiberg and Witten \cite{SW}. In the 
case ${\cal N}=4$ SYM \cite{VW}, it is related via AdS/CFT to the $SL(2,{\mathbb{Z}})$ 
invariance of type IIB string theory. The latter puts very strong constraints on the form of
the coefficients or the effective action, as it relates the string tree-level and one-loop 
terms to instanton contributions \cite{green}. The topologically twisted version of 
${\cal N}=4$ has recently been shown to realize the basic geometric Langlands 
correspondence \cite{KW}.

The question whether four-dimensional gauge theories coupled to gravity exhibit  
duality properties has a long history \cite{CJ}. Gravity itself in four dimensions is known 
to be invariant under rotations of the linearized Riemann tensor and its dual \cite{HT}. 
For toroidal compactifications of M-theory, generalized electric-magnetic dualities are 
well-known \cite{CJ,hull}. 
In this paper we will consider compactifications to asymptotically AdS$_4\times S^7$. 
Finding dualities in this case might mean significant progress towards understanding the 
physical properties of the theory describing interacting M2-branes at strong coupling.

An important source of motivation comes from \cite{Wittensl2z}, where it was argued that 
the $SL(2,{\mathbb{Z}})$ duality of abelian gauge fields in the bulk of AdS$_4$ relates 
seemingly different CFT's to each other. The CFT's in question turn out to be relevant for 
the quantum Hall effect, where the $S$-duality of \cite{Wittensl2z} had independently been 
found \cite{BD1} and  verified experimentally. In retrospect, 
this might be seen as an experimental prediction of AdS/CFT. These modular properties have 
recently been used to propose new predictions for the quantum Hall effect in graphene 
\cite{BD2}. The $S$-duality of \cite{Wittensl2z} is also related to 
the IR limit of three-dimensional mirror symmetry \cite{KS,BKW}.
It implies that electrically charged particles in one theory correspond to vortices 
in the dual.

In this paper we develop the holographic map between $SL(2,{\mathbb{Z}})$ transformations
in the bulk and dualities between three-dimensional effective CFT's away from the 
conformal fixed point. The massive deformations are induced by generalized boundary conditions in
the bulk of AdS. We show that under RG flow towards the IR, the IR theory is described by a dimension
2 current which is $S$-dual to the UV current. The UV current-current correlator is given in terms of the dual 
gauge field. This is in agreement with the field 
theory predictions of \cite{tassos}.

Before we describe our results in more detail, let us give our third piece of motivation. It
comes from the fact that conformal theories with instanton solutions in AdS have rather 
special properties. One can often write an effective boundary action for them, 
whose classical solutions reproduce the bulk instantons. The exact quantum effective action can be computed from the 
bulk at leading order in $N$. The fact that one is dealing with instantons implies that the stress-energy tensor is zero and there is no back reaction. In \cite{sebastassos} it was shown in the context of a toy model that 
for a conformally coupled scalar field with a quartic interaction potential the classical 
bulk solutions can be matched exactly with classical solutions of a boundary effective 
field theory. The boundary field theory had appeared earlier in \cite{HH}. The conformally
coupled scalar field was embedded in M-theory in \cite{dHPP}, where it was also shown that
the boundary effective action can be exactly computed using AdS/CFT.

In this paper we continue this program of conformal holography \cite{sebastassos}  for the
case of abelian gauge fields. Again, the boundary effective action can be exactly computed 
for the class of solutions under study. The large $N$ effective boundary theory which 
reproduces the bulk instantons is the self-dual topologically massive three-dimensional theory of Pilch, Townsend and van Nieuwenhuizen \cite{PTvN,DJ}. Conformal holography is expected to play a role for higher spins also.

We now summarize our main results. At the level of classical bulk solutions, we will give
the most general solution for a $U(1)$ gauge field in pure AdS$_4$, and compute the 
leading back-reaction effects in the Fefferman-Graham expansion. The on-set of back-reaction is one order up 
from the Fefferman-Graham ambiguity, and for the purpose of computing two-point functions the 
effective action can be computed by the method of holographic renormalization \cite{dHSS} without using 
back-reaction. 
We also obtain the regularity condition for Euclidean gauge 
fields, which does not seem to have appeared in the literature. This regularity condition relates
the boundary value of the electric field to the gauge invariant part of the gauge field.
In momentum space it takes the form
\be\label{1}
f_i(p)=-|p|\,A_i^{\sm{T}}(p)~,
\ee
where $f_i(p)$ is the boundary value of the electric field (see notation in the next subsection).
We explain the physical meaning of this boundary condition in Lorentzian signature: it is the
condition that waves travel from the boundary into the bulk. In terms of the original M2-geometry,
this is the condition that all matter falls into the M2-brane. We will show that self-dual
bulk solutions are in one-to-one correspondence with a choice of what we call ``self-dual
boundary conditions'':
\be\label{2}
f_i(x)=\pm \e_{ijk}\pa_jA_k(x)
\ee
for (anti-)SD solutions, respectively. Combining \eq{1} (or a massive version of it) and
\eq{2} gives the self-dual topogically massive theory of \cite{PTvN,DJ}.
We show that, 
although in flat space there are no abelian instantons, this is not necessarily true in
AdS: configurations such as \eq{1}-\eq{2} can lead to finite Euclidean action. The theory of
\cite{PTvN,DJ} has a single massive degree of freedom, and the on-shell value of the action
is explicitly written in terms of it.

The bulk theory is electric-magnetic invariant only up to boundary terms. We will compute
these explicitly for various choices of boundary conditions. Boundary conditions are
imposed by the addition of boundary terms in the action, corresponding to deformations of
the CFT. We will consider marginal operators and operators of classical dimension 4, and
show how electric-magnetic duality maps different operators into each other. We identify
configurations which are self-dual under duality, and these include the self-dual configurations
of \cite{PTvN,DJ}.

Scalars in the particular range of masses  $-d^2/4<m^2<-d^2/4+1$ in AdS$_{d+1}$ are well-known to admit two kinds of 
normalizable solutions, which give rise to two possible quantization schemes and hence two different
boundary interpretations, one in terms of an operator of dimension $\D_+$ and the other in
terms of an operator of dimension $\D_-$. In the conformal case we will be interested in, and in $d=3$,
the dimensions are $\D_+=2$ and $\D_-=1$. Four-dimensional gauge fields have the same property: both modes are 
normalizable, and two quantization schemes are available, corresponding to Dirichlet or Neumann boundary conditions. 
For Dirichlet boundary conditions where the boundary value of the gauge field (or, equivalently, the magnetic
field) is held fixed, the holographic interpretation is the usual one: $A$ is a source for a global
symmetry current in the CFT. In the Neumann quantization scheme, the electric field is held fixed instead.
The dual operator has dimension $\Delta_-=1$ and is identified with a gauge field that corresponds to the 
boundary value of the gauge field $A$. This does not present a problem because one can always construct from it
a conserved current of dimension 2 which saturates the unitarity bound $\D\geq2$. Observables of the theory are then correlation functions of the
new current. It is dual to the boundary value of the bulk magnetic field. Witten showed 
\cite{Wittensl2z} that the bulk $S$-duality operation induces an $S$-duality operation on the CFT that consists of 
first gauging the global $U(1)$ symmetry, promoting $A$ to a dynamical field, and coupling it to an external field 
$v$ via a Chern-Simons coupling. This corresponds precisely to the above discussion of boundary conditions. A third choice of boundary condition is possible in this case, a mixed one, which 
corresponds to a multi-trace deformation of the CFT. In this paper we will examine $S$-duality in the context of 
massive deformations of the CFT which induce a renormalization group flow. We will find an interesting 
interpretation of these dualities in 
terms of bulk sources. If we use the usual bulk magnetic source (one that couples to the electric field), the 
correlators that we find are the usual ones. On the other hand, if we use a bulk electric source (one 
that couples to the gauge field), we find the $S$-dual correlators.  As we mentioned, RG flow
interchanges these two. We also give a bulk proof of Witten's CFT argument that $S$-duality acts
as a Legendre transform \cite{Wittensl2z} (this was first shown in \cite{HUY}, where marginal deformations were also discussed). In fact, there are two types of Legendre transforms,
one that changes the dimension of the operator and one that doesn't. We interpret these dualities as the
particle-vortex duality of \cite{BD1}. The generic IR effective action that we find is the
one of QED in three dimensions at large $N_f$. Although this basically follows from conformal invariance,
the fact that the qualitative behavior under massive deformations is similar indicates that
QED may indeed play a role in the IR for the $U(1)$ gauge fields studied here. The supergravity
configuration that we study is ${\cal N}=2$ supergravity in four dimensions, which is 
obtained as an M-theory compactification on an $S^7$ where all scalar fields have been 
projected out. The still unknown dual SCFT is the strong-coupling limit of 2+1 dimensional SYM theory
in the large $N$ limit. We compute its effective action, and a relation with the 
large $N_f$ limit of QED is suggested. Three-dimensional $S$-duality of $U(1)$ gauge theories including Chern-Simons terms has been discussed earlier in a somewhat different context in \cite{KOO}.

The plan of the paper is as follows. In section 2 we review the discussion of scalar fields
in the special range of masses mentioned above. We will use many of those properties
in the gauge field case as well. In that section we also give the basic classical solutions
that we will use and apply the method of holographic renormalization to compute the effective
action. In section 3 we discuss electric-magnetic duality in AdS$_4$. We give a novel proof of
$S$-duality based on the first order formalism and show how the boundary terms change under
duality. Section 4 is devoted to the deformations of the boundary conditions and the deformations
of the CFT that they induce. In section 5 we show how electric-magnetic transformations
act on the general boundary conditions. In section 6 we discuss the interplay of $S$-duality,
Legendre transforms, and RG-flow. The results are summarized in figure \ref{diagram} on
page 37. In that section we compare to the known CFT results and comment on the non-abelian
generalization. Some conclusions and future lines of investigation are presented in section
7. In five appendices we present details concerning $S$-duality, the various solutions, holographic renormalization,
and supersymmetry of the solutions.

\subsection*{Notation and conventions}\label{notation}

$A_i(r,x)$ will denote the bulk gauge field and $A_i(x)=A_{(0)i}(x)$ its restriction to the boundary at $r=0$. The indices $i=0,1,2$ run over the boundary. The electric field will be denoted by $E_i(r,x)=F_{ri}(r,x)$ and its boundary component is $f_i(x)=E_i(0,x)$. We will use the following field redefinitions by a $\th$-term:
\bea
{\sf{E}}&=&E-\th\,F\nn
{\sf{f}}&=&f-\th\,F\nn
{\sf{v}}&=&v-\th\,A~.
\eea
We will use form notation for three-dimensional quantities, and will use a short-hand notation for contractions: $Af=A_if^i$ etc. Boundary indices of $r$-dependent quantities are raised and lowered with $g_{ij}(r,x)$, indices of quantities evaluated at $r=0$ are raised and lowered with $g_{(0)ij}$. We will use a projector onto the part of the gauge field transverse to the momentum:
\be\label{id2}
\Pi_{ij}=g_{ij(0)}-{\pa_i\pa_j\over\Box}~,~A^{\sm{T}}_i=\Pi_{ij}A_j=(\Pi A)_i~.
\ee
We will denote $M=\Box^{1/2}=\sqrt{|\Box|}$ which both in Lorentzian and Euclidean signature is a positive definite quantity ($M=|p|$ in momentum space). The quantity $t$ will be +1 in Euclidean signature, and -1 in Lorentzian signature.

The boundary magnetic field will be denoted by a vector $F$,
\be
F_i=\half\e_{ijk}F_{jk}=\e_{ijk}\pa_j A_k~.
\ee

Twiddles denote Legendre transforms, tildes denote $S$-duality. So $\ti S$ is the Legendre transformed action, $S'$ denotes its $S$-dual.

Boundary operators will be denoted ${\cal O}_{\D i}(x)$, with $\D=1$ for gauge fields and $\D=2$ for currents.
Boundary sources are denoted by $J_i(x)$.

\section{The bulk solution}

\subsection{Duality for scalar fields}\label{scalardualities}

In this section we review some facts known for scalar fields in AdS$_4$ which will set the stage of the discussion of gauge fields in the rest of the paper. It will also serve to fix notation.

Scalar fields in AdS$_{d+1}$ with mass within the range
\be\label{massrange}
-{d^2\over4}<m^2<-{d^2\over4}+1
\ee
are well-known to admit two types of possible boundary conditions at the boundary preserving the asymptotic symmetry group and leading to finite energy \cite{isham,bf,Breitenlohner:1982jf}. For definiteness, consider the Fefferman-Graham form of the metric
\be\label{FG}
\dd s^2={\ell^2\over r^2}\left(\dd r^2+g_{ij}(r,x)\,\dd x^i\dd x^j\right)
\ee
where $g_{ij}(r,x)$ has an expansion in the distance $r$ to the boundary, such that at the boundary $r=0$ the metric is the Poincare metric of AdS. A scalar field of mass $m$ will now have the asymptotic expansion
\be
\f=r^{\D_-}\f_{(0)}(x)+\ldots+r^{\D_+}\f_{(2\D-d)}(x)+\ldots
\ee
where $\D_\pm={d\over2}\pm\sqrt{{d^2\over4}+m^2}$. $\f_{(0)}$ and $\f_{(2\D-d)}$ are not determined by the field equations. They are boundary conditions and have a holographic CFT interpretation as the source and 1-point function of an operator of dimension $\D_+$, respectively, to which the source is coupled. For the range of masses \eq{massrange}, it was argued \cite{klebanovwitten} that the two quantization schemes,
\bea\label{scDN}
\d\f_{(0)}(x)&=&0~~~~~~{\mbox{Dirichlet}}\nn
\d\f_{(2\D-d)}(x)&=&0~~~~~~{\mbox{Neumann}}~,
\eea
correspond to two possible boundary CFT's. The usual CFT is the one discussed above, where a source $\f_{(0)}$ couples to an operator of dimension $\D=\D_+$. The usual AdS/CFT recipe can then be summarized as (up to terms that contribute contact terms in the two-point functions and which vanish identically in our case):
\bea\label{recipe}
\f_{(0)}&=&J(x)\nn
S_{\sm{on-shell}}[\f_{(0)}]&=&-W[J]\nn
\bra {\cal O}_{\D}(x)\ket_J&=&-{\d W\over\d J}={\d S_{\sm{on-shell}}\over\d\f_{(0)}(x)}=(2\D-d)\,\f_{(2\D-d)}\nn
\bra {\cal O}_{\D}(x){\cal O}_{\D}(x')\ket_J&=&-{\d^2 S_{\sm{on-shell}}\over\d\f_{(0)}(x)\d\f_{(0)}(x')}+\bra {\cal O}_{\D}(x)\ket\bra {\cal O}_{\D}(x')\ket~,
\eea
and likewise for the higher-point functions. The signature here is Euclidean. Our conventions are such that in the Lorentzian we have $S_{\sm{on-shell}}=W$ and hence $\bra {\cal O}_{\D}\ket=-(2\D-d)\f_{(2\D-d)}$.

Thus, the renormalized bulk on-shell action is interpreted as minus the (renormalized) generating functional of connected correlation functions in the CFT, $W$, at leading order in $N$. We can also define the effective action of the boundary CFT, which is its Legendre transform:
\be
\Gamma[\f_+]=-\int\dd^dx\,\phi_-(x)\phi_+(x) +W[\f_-]~.
\ee

The second CFT is what we call the ``dual CFT''  \cite{klebanovwitten}, and it contains an operator of dimension $\D_-$, which couples to a source of dimension $\D_+$. This theory is defined as the Legendre transform of the usual CFT,
\be
W[\f_-]=\ti W[\f_+]+\int\dd^dx\,\f_-\f_+~.
\ee
Its effective action $\ti\Gamma$ is given by
\be
\ti W[\f_+]=-\ti\G[\f']+\int\dd^dx\,\f_+\f'~.
\ee
Comparing both equations, we see that the effective action of the dual CFT is identified with the bulk action, for $\f'=-\f_-$, and the generating functional of the dual CFT is identified with the effective action of the usual CFT:
\be
\ti\G[-\f_-]=-W[\f_-]=S[\f_-]~,~\ti W[\f_+]=\G[\f_+]~.
\ee
where $\f_-$ is interpreted as the expectation value of an operator of dimension $\D_-$ and $\ti J=\f_+$ is the dual source. The operator\footnote{For the conformal equivalence between $\f_+$ and $\f_-$, see \cite{Dobrev}.} is given by:
\be
\bra{\cal O}_{\D_-}\ket_{\ti J}={\d\ti W[\f_+]\over\d\f_+}~.
\ee
The fact that we are in the mass range \eq{massrange} ensures that $\D_-$ is above the unitarity bound $\D_->{d-2\over2}$. 

For spin-1 fields we have $\D_\pm={d\over2}\pm\sqrt{{(d-2)^2\over4}+m^2}$, hence for massless spin-1 fields in AdS$_4$ we get $\D_-=1$, $\D_+=2$ as in the case of conformally coupled scalar fields. As we will see, this fact implies that massless gauge fields and conformally coupled scalar fields share many important properties. For scalars in the range of masses \eq{massrange}, a generalized choice of boundary conditions is possible where the total energy is finite and conserved \cite{HM,designer}:
\be\label{genbc}
\f_+=\f_+(\f_-)={\cal F}'(\f_-)
\ee
for some function ${\cal F}'$. Since $\f_-$ is itself a gauge invariant operator in the CFT, \eq{genbc} deforms the boundary theory by a multiple-trace operator \cite{wittendt}. For example, in the case at hand, in order to obtain a generalized boundary condition that preserves all of the AdS isometries,
\be\label{trtr}
\f_+=\a\,\f_-^2
\ee
we need to deform the CFT by a marginal triple-trace operator $\int\dd^3x\,\f_-^3(x)$.

In order to allow for generalized boundary conditions \eq{genbc}, the variational principle needs to be modified. Indeed, the usual variational principle implies either Dirichlet or Neumann boundary 
conditions, which correspond to the $\a=\infty$ and $\a=0$ limits of \eq{trtr}, respectively. In  \cite{MSS}, and 
independently in \cite{designer}, a recipe was given where the variational principle is well-defined and 
automatically leads to boundary condition \eq{genbc}. This amounts to adding a boundary term to the action such that 
the boundary condition is enforced. Let us see this in some detail. Under usual Dirichlet/Neumann boundary conditions 
\eq{scDN}, the bulk path integral computes the CFT correlator with a {\it fixed} external source,
\be
Z[J]=\int_{\f_{\mp}=J}{\cal D}\f\,e^{-S[\f]}=\bra e^{\int\sm{d}^dx\,J(x){\cal O}_{\D_{\pm}}(x)}\ket_{\sm{CFT}}~,
\ee
and integration over all the bulk fields should be understood here. On the other hand, the boundary condition \eq{genbc} is achieved if we integrate over both the bulk and the boundary values of the scalar field:
\be\label{pathi}
Z=\int{\cal D}\f_-\int_{\f|_{r=0}=\f_-}{\cal D}\f\,e^{-S[\f]+S_{\mbox{\tiny{bdy}}}[\f_-]}=\int{\cal D}J\,e^{S_{\sm{bdy}}[J]}\bra 
e^{\int\sm{d}^dx\,J(x){\cal O}_{\D_+}(x)}\ket_{\sm{CFT}}~.
\ee
Thus, the path integral is evaluated without any boundary conditions \cite{MSS}. In general, the saddle point approximation for 
\eq{pathi} gives two terms. The first one is the bulk equations of motion. The second one is proportional to
\be
-\int\dd^3x\,\f_+\d\f_-+\d S_{\sm{bdy}}[\f_-]
\ee
in other words, excluding the trivial case of Dirichlet boundary conditions, the general boundary condition is
\be
\f_+={\d S_{\sm{bdy}}\over\d \f_-}[\f_-]~,
\ee
again for the relevant case of the conformally coupled scalar field in AdS$_4$. In \cite{MSS} it was shown that, for 
quadratic $S_{\sm{bdy}}[\f_-]$, the modified two-point functions extracted from this recipe agree with those of the 
field theory where an explicit double-trace deformation is introduced in the action. In this paper we will use a 
similar procedure for gauge fields.

\subsubsection*{Lorentzian signature}

In Lorentzian signature there are a number of subtleties in the above picture. These have 
to do with the fact that propagating, normalizable modes now exist, and hence 
the solution is not uniquely determined by the boundary condition at the boundary plus regularity of the solution 
at infinity. In particular, the functional relation between $\f_+$ and $\f_-$ is no longer fixed. The propagating modes
correspond to a non-trivial choice of state in the boundary theory, and their wave-functions have to be included in the 
path integral \cite{don}. The solution that is analytically continued from the Euclidean corresponds to the vacuum state -- in that
case, the saddle point approximation to the bulk partition function gives the Euclidean CFT partition function, which when
analytically continued gives the vacuum amplitude. In the Lorentzian, choosing any other solution than the regular one gives
an amplitude between non-trivial states.


In this paper we will largely use the regular boundary condition. In section \ref{regularitysec} we will derive this condition for gauge fields and give a purely Lorentzian interpretation of it: instead of as a boundary condition at $\S_{\pm}$, it can be regarded as a condition for the flux to be purely incoming at the boundary. This is the natural boundary condition in the original M2-brane geometry. This implies that we can in fact impose this as an independent boundary condition
at the boundary. 

\subsection{Gauge fields: solutions and dual interpretation}

We will now solve the equations of motion for the gauge field. For later reference, we give the Lorentzian form of the action:
\be\label{actionwtheta}
S[A]=\int\dd^4x\,\sqrt{-g}\left(-{1\over4g^2}F_{\m\n}^2+{\th\over32\pi^2}\,\e^{\m\n\l\s}F_{\m\n}F_{\l\s}\right)~.
\ee
The $\th$-term is the appropriate one for a spin manifold. It gives a Chern-Simons term at level 1/2. If the manifold is not spin, the coefficient would have an additional factor of 2. The equations of motion that follow from the above matter part of the action are:
\be
\nabla^\m F_{\m\n}=0~,
\ee
In the Fefferman-Graham coordinate system \eq{FG}, we can do the following asymptotic expansion for a massless field of spin 1:
\bea
A_r(r,x)&=&A_r^{(0)}(x)+r\,A_r^{(1)}(x)+r^2\,A_r^{(2)}(x)+\ldots\nn
A_i(r,x)&=&A_i^{(0)}(x)+r\,A_i^{(1)}(x)+r^2\,A_i^{(2)}(x)+\ldots
\eea
Solving Maxwell's equations then determines all the higher coefficients $A_r^{(n)}, A_i^{(n)}$, $n\geq2$, in terms of 
the boundary conditions $A_i^{(0)}(x),A_i^{(1)},A_r^{(0)},A_r^{(1)}$, as well as the metric. The radial components 
$A_r^{(0),(1)}$ are of 
course not real degrees of freedom as they can always be gauged away. Notice that $A_i^{(1)}$ is not left totally undetermined by the equations of motion. There is a constraint on its gauge-invariant part, which for future convenience we will denote $f_i(x)$:
\be
\nabla_i^{(0)}f^i(x)=0~,
\ee
where the covariant derivative is defined with respect to the boundary metric $g_{(0)ij}$,
boundary indices are raised and lowered with respect to this metric, and where we defined
\bea
f_i(x)&=&A_i^{(1)}(x)-\pa_iA_r^{(0)}(x)\nn
F_{ri}(r,x)&=&f_i(x)+{\cal O}(r)~,
\eea
with a conventional minus sign in Lorentzian signature.
Of course, $f_i(x)$ is nothing but the boundary value of the electric field $E_i=F_{ri}$. The fact that the bulk equations only fix $f_i(x)$ up to an arbitrary conserved vector is called the Fefferman-Graham ambiguity.

Since $f_i(x)$ is conserved, in three dimensions and if $H^1(M)=0$ where $M$ is the boundary, we can introduce a vector $v_i$ such that
\be
f^i=\e^{ijk}_{(0)}\pa_jv_k~.
\ee
Again, the epsilon tensor is defined with respect to $g_{(0)ij}$.
From now on, we will write $A_i(x)$ instead of $A_i^{(0)}(x)$ for the boundary value of the 
bulk field $A_i(r,x)$.

We can take advantage of conformal invariance to completely solve the Maxwell equations 
throughout the bulk. In Appendix \ref{regularityapp} we do this for pure AdS. The general solution 
in that case is
\be\label{Aiexpansion}
A_i^{\tiny{\mbox{T}}}(r,p)=A_i^{\tiny{\mbox{T}}}(p)\cosh(|p|r)+{1\over|p|}\,f_i(p)\sinh(|p|r)~,
\ee
in momentum space. Here $A_i^{\tiny{\mbox{T}}}$ is the transverse, gauge invariant part of 
the gauge field, $A^{\tiny{\mbox{T}}}_i(p)=\Pi_{ij} A_j(p)$ with the projector defined in 
\eq{id2}. By construction, $f_i(x)$ is also gauge invariant.

In this paper we will take advantage of the holographic interpretation in the dual CFT that is relevant to the Neumann and 
mixed boundary problem, as discussed in sections 1 and 2.1. In the standard CFT, the boundary values of the gauge field 
$A$ (or alternately the magnetic field) are fixed in terms of a boundary source for a global $U(1)$ current. The 
boundary values of the electric field correspond to the conserved current. In the dual CFT, the global symmetry has 
been gauged and $A$ has been promoted to a dynamical field that is integrated over in the path integral. Its one-point 
function corresponds holographically to the boundary value of the (transverse part of the) gauge field. It gives 
rise to a new conserved current $\ti f=*\dd A$ which corresponds to the magnetic field and which is the new
gauge invariant observable of the theory saturating the unitarity bound $\D\geq 2$. Finally, a new background 
field $v$ has been introduced which couples to this conserved current. This new background field corresponds to the 
bulk electric field via $f=*\dd v$. 

\subsection{Holographic renormalization}\label{backreaction}

We now summarize the analysis of the coupled gravity-Maxwell system and the holographic renormalization of the action, which is done in Appendix \ref{holren}.

The total Euclidean action is written in \eq{Euclaction}. In the first order 
formalism, we simply replace the matter part of the action by the corresponding first order matter
action \eq{1storderform}:
\be
S=-{1\over16\pi G_N}\int\dd^4x\sqrt{g}\,\left(R-2\L\right)-{1\over8\pi G_N}\int\dd^3x\sqrt{\g}\,K+S_{\sm{bulk}}[A,E]+S_{\sm{bdy}}[A,E]~.
\ee
The components of the stress-energy tensor are
\bea\label{stressenergy}
T_{\m\n}&=&{r^2\over\ell^2}\,\t_{\m\n}~;\nn
\t_{rr}&=&\half(E^2-F^2)~,\nn
\t_{ri}&=&E^jF_{ij}=\e_{ijk}E_jF_k~,\nn
\t_{ij}&=&E_iE_j-F_iF_j-\half\,g_{ij}(E^2-F^2)~.
\eea
The stress-energy tensor is manifestly invariant under  electric-magnetic transformations (which are hyperbolic in Euclidean signature), as it should. In this formulation, it is obvious that the stress-energy tensor is zero if and only if
\be
E_i=\ve\, F_i~,
\ee
where $\ve^2=1$, i.e. the solution is (anti-)self-dual.

In Appendix \ref{holren}, Einstein's equations are solved in the coordinate system \eq{FG}:
\bea\label{grx}
g(r,x)&=&g_{(0)}+r^3g_{(3)}+r^4g_{(4)}+\cdots\nn
\nabla^jg_{(3)ij}&=&-{16\pi G_N\over3\ell^2}\,f^jF_{ij}\nn
g_{(4)ij}&=&-{4\pi G_N\over\ell^2}\left(f_if_j-F_iF_j-{1\over4}\,g_{ij}(f^2-F^2)\right)~,
\eea
where we have chosen a Ricci-flat boundary metric, hence $g_{(2)}=0$. $g_{(3)}$ is otherwise undetermined by the 
field equations and has the interpretation as the stress-energy tensor of the boundary theory. This is the usual 
Fefferman-Graham ambiguity for the metric and it is fixed once one imposes regularity of the bulk solution. As 
explained in the Appendix, the middle equation above is a Ward identity that follows from the diffeomorphism 
invariance of the on-shell action. In fact, it is of the form 
\be
\nabla^j\bra T_{ij}\ket=T_{ir}=-F_{(0)ij}f^j~.
\ee
$\bra T_{ij}\ket$ is also the quasi-local stress-energy tensor of Brown and York \cite{BY}. The above identity measures the matter flow through the boundary. From \eq{stressenergy} we see that it vanishes if $f_i$ is proportional to $F_i$, which are indeed AdS-preserving boundary conditions. Using the fact that the boundary current is
\be
\bra J_i\ket=-f_i~,
\ee
the above is the expected field theory Ward identity \cite{dHSS,BFS}.

In Appendix \ref{holren} we show that, after subtracting the divergent part of the effective action by adding the 
usual counterterm, there is no near-boundary contribution to the on-shell action coming from the gravity part of the 
action. 
Taking into account back-reaction would result in 
triple and higher order terms in the boundary operators, which we may neglect since we are only interested in
two-point functions. Therefore, in this approximation the only contribution is from the matter part of the action, 
which we analyze in the next subsection.

\subsection{Regularity and incoming boundary conditions}\label{regularitysec}

We now impose the usual normalizability conditions in AdS \cite{isham,bf,vijayper}. The situation for gauge fields is similar to the case of conformally coupled scalar fields. Let us denote the boundary momentum by $p=(\o,k)$. In Euclidean signature, as well as for Lorentzian tachyonic modes $\o^2<k^2$, the general solution of the wave equation is
\be
A_i^{\tiny{\mbox{T}}}(r,p)=\half\left(A_i^{\tiny{\mbox{T}}}(p)+{1\over|p|}\,f_i(p)\right)e^{|p|r}+\half\left(A_i^{\tiny{\mbox{T}}}(p)-{1\over|p|}\,f_i(p)\right)e^{-|p|r}~.
\ee
The first solution blows up at $r=\infty$. It is clear that if we demand 
\be\label{regularity}
A_i^{\tiny{\mbox{T}}}(p)+{1\over|p|}\,f_i(p)=0~,
\ee
then regularity is ensured. One can check \cite{marolf} that the solutions are normalizable. In position space the above is a non-local relation:
\be
A_i^{\tiny{\mbox{T}}}(x)=-{i\over \Box^{1/2}}\,f_i(x)=-\int\dd^3y\,{f_i(y)\over|x-y|^4}~.
\ee
Notice that this is simply the linear boundary condition \eq{dtbc} where $m=-i\Box^{1/2}$ is an operator. 
For later convenience we re-write the gauge invariant regularity condition as 
\be\label{gaugeinvreg}
f=-\sqrt{-\Box}\Pi A~.
\ee 

For regular bulk solutions, the gauge field and field strength now take the following form:
\bea\label{regulargauge}
A_i(r,x)&=&A_i(x)\,e^{-|p|r}\nn
F_{ri}(r,p)&=&-|p|\Pi_{ij}A_j(p)\,e^{-|p|r}\nn
F_{ij}(r,p)&=&i(p_iA_j(p)-p_jA_i(p))\,e^{-|p|r}\nn
F_i(r,p)&=&i\e_{ijk}p_jA_k(p)\,e^{-|p|r}~.
\eea

Using the equations of motion, it is also easy to see that if we impose \eq{regularity} at $r=0$ then we also have
\be\label{regularityr}
A_i^{\sm{T}}(r,p)+{1\over|p|}\,E_i(r,p)=0
\ee
for all $r$. This will ensure that the contribution from $r=\infty$ to the on-shell action vanishes. The transverse 
part of the gauge field is defined in \eq{transverse}, hence the above equation is gauge invariant with respect to 
boundary gauge transformations of the dual CFT (see end of section 2.2).

In Lorentzian signature, all the modes satisfying $\o^2>k^2$ are normalizable and so there is no regularity condition for these modes. There is, however, a natural choice of boundary condition \cite{herzogson} which corresponds to imposing that all waves are ingoing at the boundary -- that is, all waves travel from the boundary into the bulk. Notice that this is true both at $r=0$ and $r=\infty$; since we are dealing with pure AdS, waves that are ingoing at the boundary remain ingoing near $r=\infty$. From the field theory point of view, this choice is natural because it corresponds to defining the vacuum in the Lorentzian field theory as the analytically continued Euclidean vacuum; with this choice of boundary conditions, a Lorentzian transition function in the vacuum corresponds to the Euclidean partition function. This is the choice that leads to {\it absorption} cross section by the M2-brane. Let us explain this point. When solving the wave equation in the original M2-brane geometry \cite{maldacena}, 
\bea\label{M2}
\dd s^2&=&f^{-2/3}\,(-\dd t^2+\dd\vec x^2)+f^{1/3}(\dd R^2+R^2\dd\Omega_7^2)\nn
f&=&1+{2^5\pi^2N\ell_{\sm{Pl}}^6\over R^6}~,
\eea
with a non-trivial four-form flux, 
a choice of {\it incoming} waves must be made at the horizon -- in other words, only the modes describing infalling matter are kept \cite{unruh,GK} (see \cite{GKT} for the higher-spin extension). Taking the near-horizon limit of \eq{M2} we get
\be
\dd s^2={\ell^2\over r^2}(\dd r^2-\dd t^2+\dd\vec x^2)+4\ell^2\dd\Omega_7^2
\ee
with $2\ell=\ell_{\sm{Pl}}/(2^5\pi^2N)^{1/6}$ where $R\sim 1/r$. Hence, the modes absorbed by the M2-brane will be modes travelling from the boundary towards the bulk. Such modes have the form $e^{-i\o t+i|p|r}$: indeed, for such modes, as time elapses the wave front increases $r$ and so they propagate to the bulk. Since we are doing classical field theory near the M2-brane, it is natural to take the waves to be ingoing at the horizon. Using the change of coordinates in \cite{herzogson}, it is easy to check that this remains true in the non-extremal case. We will now see this for the gauge field.

We write the general form of the solution of Maxwell's equations \eq{Aiexpansion} as
\bea\label{Lor}
A_i^{\tiny{\mbox{T}}}(r,p)&=&A_i^{\tiny{\mbox{T}}}(p)\,\cos(|p|r)+{1\over|p|}\,A_{i(1)}(p)\,\sin(|p|r)\nn
&=&\half\left(A_i^{\tiny{\mbox{T}}}(p)+{1\over i|p|}\,A_{i(1)}(p)\right)e^{i|p|r}+ \half\left(A_i^{\tiny{\mbox{T}}}(p)-{1\over i|p|}\,A_{i(1)}(p)\right)e^{-i|p|r}
\eea
where now $|p|=\sqrt{\o^2-k^2}$. Analytic continuation from the Euclidean is done by $|p_{\sm{E}}|\rightarrow -i|p_{\sm{L}}|$, therefore the regularity condition \eq{regularity} turns into
\be\label{Lreg}
A_{(0)}^{\sm{T}}(p)+{1\over i|p|}\,A_{(1)}(p)=0
\ee
whenever $\o>0$. This leaves only the first factor in \eq{Lor}. Taking into account the fact that $A_{(0)}$ contains a factor of $e^{-i\o t}$, we get a wave $e^{-i\o t+i|p|r}$. This agrees with the behavior of the natural boundary condition in \cite{herzogson}. 

One can check that, by construction, the bulk fields are purely negative frequency in the far past and purely positive frequency in the far future, as one would expect from the analytic continuation of the Euclidean 2-point function which is the Feynmann Green's function.

\subsubsection*{In-going vacuum and vanishing action}

Incoming boundary conditions of the type just found in the Lorentzian describe a special vacuum that we will now analyze in more detail. As we will see, they have exactly zero action. Expand the bulk solutions again in plane waves:
\bea\label{Arx}
A(r,x)&=&\int\dd^3p\,a(p)\,e^{i|p|r+ip\cdot x}+\mbox{c.c.}\nn
&=&\int_0^\infty\dd\o\int\dd^2k\left(a(p)\,e^{i(|p|r-\o t)}+a^*(-p)e^{-i(|p|r+\o t)}\right)e^{ikx}+\mbox{c.c.}
\eea
where $k$ is a 2-vector that labels spatial momenta along the boundary.
As before, the in-going vacuum requires vanishing of the second factor, therefore
\be\label{ing}
a(p)=a(\o,k)=0~~~\mbox{if}~\o<0~.
\ee
Notice that only the sign of $\o$ matters here, with respect to which positive frequencies are defined. We now look at the boundary expansion of \eq{Arx}:
\be
A(r,x)=\int_0^\infty\dd\o\int\dd^2k\left[A_{(0)}(p)\cos(|p|r)+iB_{(1)}(p)\sin(|p|r)\right]e^{-i\o t+ikx}+\mbox{c.c.}
\ee
where $A_{(0)}(p)=a(p)+a^*(-p)$, $B_{(1)}(p)=a(p)-a^*(-p)$, and $A_{(0)}(p)$ defines the natural field theory vacuum 
$A_{(0)}(p)|0\ket=0$, since at $r=0$ $B_{(1)}$ disappears. The in-going vacuum \eq{ing} gives
$A_{(0)}(p)=\mbox{sgn}(\o)\,B_{(1)}(p)$. Defining the vacuum in terms of $A_{(0)}$ makes sense in the dual CFT, where 
$A_{(0)}$ is interpreted as an operator.

A similar analysis for the electric field gives:
\be
E(r,x)=\int_0^\infty\dd\o\int\dd^2k\left[G_{(1)}(p)\sin(|p|r)+f(p)\cos(|p|r)\right]e^{-i\o t+ikx}+\mbox{c.c.}
\ee
where $f(p)=-i|p|B_{(1)}(p)$ and $G_{(1)}(p)=|p|A_{(0)}(p)$.
The in-going boundary condition now gives
\be\label{nreg}
f(p)=-i|p|\,\mbox{sgn}(\o)\,A_{(0)}(p)
\ee
which is the generalization of \eq{Lreg} to arbitrary frequencies.

It is now straightforward to check that for such states the bulk action is zero. As we will 
prove in the next section, the on-shell action is proportional to $\int\dd^3x\,fA$.
Expanding this in Fourier modes, and using \eq{nreg}, this is readily seen to be identically zero. On the other hand, the Hamiltonian is non-zero.

\subsection{Self-duality and topologically massive theories}\label{sdtmt}

We will now analyze self-dual solutions of the Euclidean field equations. The Euclidean action is:
\be
S={1\over4g^2}\int\dd^4x\sqrt{g}\,F_{\m\n}F^{\m\n}+{\th\over32\pi^2}\int\dd^4x\,\sqrt{g}\,\e^{\m\n\l\s}F_{\m\n} F_{\l\s}~.
\ee
We will prove three main results concerning self-dual solutions. First, we will show that there is a one-to-one correspondence between self-dual solutions in the bulk and a choice of boundary conditions that we will call ``self-dual boundary conditions''. Second, we will show that bulk self-dual solutions together with a boundary condition that breaks conformal invariance leads to the self-dual topological theory of \cite{PTvN}. Finally, we will analyze regularity of the solutions and show that massive solutions are regular for negative value of the deformation parameter. 

\subsubsection*{Self-duality and topologically massive theories}

We start with (anti-)self-dual solutions in the bulk. These satisfy:
\be\label{sd}
F_{\m\n}=-{\ve\over2}\,\e_{\m\n\a\b}F^{\a\b}~,
\ee
This equation is conformally invariant, therefore in the Poincare patch this is an equation on ${\mathbb R}_+^4$. $\ve$ can take the values $\pm1,0$. We get:
\be\label{selfduality}
F_{ri}=-{\ve\over2}\,\e_{ijk}F_{jk}~~~,~~~F_{ij}=-\ve\,\e_{ijk}F_{rk}~.
\ee
The condition \eq{sd} is a first order equation. Together with the Bianchi it implies the field equation. Therefore it determines one of the two boundary conditions. From \eq{selfduality}:
\be\label{sdbc}
f_i=-\ve\,\e_{ijk}\pa_jA_k~.
\ee
This condition is what we will call the ``self-dual boundary condition''.

Next we will show that, given the self-dual boundary condition \eq{sdbc}, any solution of the bulk equations is self-dual 
and hence satisfies \eq{selfduality} throughout the whole of AdS. This is easy to see. We use the solution of the equations
of motion \eq{Aiexpansion} to work out the field strength,
\bea
F_{ri}(r,p)&=&|p|\,A_i^{\tiny{\mbox{T}}}(p)\,\sinh(|p|r)+f_i(p)\,\cosh(|p|r)\nn
F_{ij}(r,p)&=&i(p_iA_j-p_iA_j)\,\cosh(|p|r)+{i\over|p|}\,(p_if_j-p_jf_i)\,\sinh(|p|r)~.
\eea
Filling the boundary condition \eq{sdbc}, it is easy to see that the self-duality condition \eq{selfduality} is satisfied


This also implies that the self-dual boundary condition \eq{sdbc} is enough to ensure everywhere vanishing of the 
stress-energy tensor (see section \ref{backreaction}).

Now we will consider massive deformations of the boundary theory. This is analogous to the scalar field case, where a 
choice of massive boundary condition corresponds to adding relevant terms to the boundary theory. In this case, the relevant massive
boundary condition is:
\be\label{dtbc}
A_i^{\tiny{\mbox{T}}}={1\over m}\,f_i
\ee
where $m$ is a dimensionful parameter. We will call this the ``massive boundary condition''.

Combining \eq{dtbc} with \eq{sdbc}, we see that the boundary value of the gauge field satisfies the following equation:
\be\label{SD}
A_i^{\tiny{\mbox{T}}}=-{\ve\over m}\,\e_{ijk}\pa_jA_k^{\tiny{\mbox{T}}}.
\ee
Surprisingly, this is the self-dual abelian theory in three dimensions of \cite{PTvN}! Deser and Jackiw showed \cite{DJ} that this theory is equivalent to topologically massive QED in three dimensions. We will come back to its interpretation later.

\subsubsection*{Regularity and massive solutions}

Finally, we analyze regularity of the different solutions discussed here. Regularity relates the two boundary conditions of the solution in a non-local way, in this case $A_i(x)$ and $f_i(x)$, as we showed explicitly in \eq{regularity}. 
We will first construct a regular self-dual solution, that is we combine equation \eq{regularity} with \eq{sdbc}. We get the following non-local equation for the boundary gauge field:
\be\label{rmas}
A_i^{\tiny{\mbox{T}}}=-{i\ve\over\Box^{1/2}}\,\e_{ijk}\pa_jA_k^{\tiny{\mbox{T}}}~,
\ee
again a gauge invariant equation. This gives the most general regular self-dual solution in the bulk. We solve this equation in Appendix \ref{solutionSD}. In fact, this is the most general solution with vanishing stress-energy tensor.

Next we construct regular massive solutions. Thus we combine \eq{dtbc} and \eq{regularity}. This gives the following condition in momentum space:
\be\label{sign(m)}
|p|=-m~.
\ee
Notice that in this case regularity does not act as a boundary condition on the boundary values of the fields. Instead, it puts the momentum of the boundary theory on the mass shell. Obviously, this is only possible if $m<0$. This means that, for regular solutions, the deformation parameter of the boundary theory has negative sign. Thus, in this case we can still impose a second boundary condition. It is indeed possible to combine the self-dual boundary condition \eq{sdbc} with the massive one \eq{dtbc}, subject to the regularity condition \eq{sign(m)}. 

This fact is actually generic: the deformation parameter in the linear boundary condition, which corresponds to the deformation parameter of the double trace deformation, is required to be negative in order for two-point functions of dual operators to be positive definite. This is reminiscent of the regularity analysis in \cite{dHPP} for triple-trace deformations and the constraints on the deformation parameters found there. 

\subsection{Abelian instantons in AdS$_4$}\label{iads4}

We have obtained regular self-dual solutions of the Euclidean equations of motion. We can now ask whether these solutions actually have finite action and correspond to instantons. In flat space there are of course no abelian instantons because one cannot form a topological number. However, things may change in AdS due to the boundary terms.

The on-shell value of the action is computed in the next section:
\be
S_{\sm{on-shell}}=-\int\dd^3p\,A(p)f(-p)~,
\ee
where we have also included the contribution of explicit boundary terms (worked out in \eq{fF}).
We will now ask whether there are solutions such that this number is finite. We consider the regular instanton solutions of section \ref{sdtmt}. The explicit solution of \eq{rmas} is given in \eq{Asolution} of the Appendix. First of all we remark that the action simply reduces to the Chern-Simons term $\int A\wedge\dd A$, which for these solutions can be written as:
\be
S_{\sm{on-shell}}={1\over g^2}\int\dd^3p\,|p|A_i^{\tiny{\mbox{T}}}(p)A_i^{\tiny{\mbox{T}}}(-p)~.
\ee
We should remember the constraints \eq{B3}. In the notation in the Appendix,
\be
A_i^{\tiny{\mbox{T}}}(p)A_i^{\tiny{\mbox{T}}}(-p)=-\left(1+{4p_1^2p_2^2\over(p_1^2+p_2^2)^2}\right)f(p)f(-p)~.
\ee
Redefine now
\be
F(p)=\sqrt{1+{4p_1^2p_2^2\over(p_1^2+p_2^2)^2}}\,f(p)~.
\ee
The action becomes
\be
S_{\sm{on-shell}}={2\sqrt{2}\over g^2}\int\dd p_1\int\dd p_2\,\sqrt{p_1^2+p_2^2}\,|F(p)|^2
\ee
and the integral is over the remaining ${\mathbb{R}}^2$. The following choice will obviously make the action finite:
\be\label{Fp}
F(p)={2^{1/4}\sqrt{\pi a\l}\over a^2+p_1^2+p_2^2}~.
\ee
$\l$ is an arbitrary dimensionless constant. $a$ is an arbitrary scale and as usual for instantons the value of the action is independent of it. We get
\be
S_{\sm{on-shell}}={8\pi^2\over g^2}\,|\l|~.
\ee
The value of $|\l|$ can be fixed by fixing, for example, the holonomy of the gauge field. Since $\l$ must be independent of the coupling, it is remarkable that the action is in fact proportional to $1/g^2$. 

\section{$S$-duality in AdS$_4$}

We will now analyze the action of $S$-duality on the bulk action. Since electric-magnetic rotations are not canonical transformations, they act non-trivially on the canonically conjugated variables $A$ and $E$. It is well-known that, whereas electric-magnetic transformations are a manifest symmetry of the equations of motion, they are not a manifest symmetry of the action. They have to be realized at the level of the gauge potential, and this is non-trivial. The solution of this problem is well known (see for example \cite{DT,dght,ss,Deser:2004xt}). Here we use the first order formulation in \cite{deser,Deser82}.

In AdS, electric-magnetic invariance is broken due to the presence of the boundary. Therefore, an electric-magnetic transformation changes the boundary terms in the action. We will first explicitly show the $S$-duality of the bulk part of the action and compute the boundary terms. We will then generalize this to the full $SL(2,{\mathbb{R}})$ ($SL(1,1)$ in the Euclidean case). Witten's proof of $S$-duality for abelian theories 
involves introducing an additional two-form field in the action, together with an enlarged 
gauge symmetry \cite{wittenabelian}. $S$-duality is then showed after integrating out two different fields. We 
will instead give two elementary proofs that rely only on the first order formalism of the 
action. The advantage of our first proof is that it does not involve integrating out fields 
but only a simple field redefinition. The price we pay is that Lorentz invariance is not 
manifest. In this proof, we write the action in terms of first order fields $\sf E$ and $A$, 
where $\sf E$ is the redefined conjugate momentum. In terms of these variables, the action 
is manifestly invariant under $S$-duality including the $\th$-angle. The second proof, given in Appendix \ref{2nds},
involves integrating out either $E$ or $A$, to get two actions with the same form which are 
dual to each other. It presents some similarities to the methods in \cite{wittenabelian,yolanda}.

Then, using holographic renormalization, we will show how $S$-duality acts on the boundary effective potential, and on the boundary 1- and 2-point functions, and how the Legendre transform arises.

$T$-transformations simply act by $\th\rightarrow\th+2\pi$ in \eq{actionwtheta}. These do not leave the action invariant but they do leave the partition function 
invariant.

\subsection{The first order formalism}\label{1of}

We first work out the first order form of the action \eq{actionwtheta} in Lorentzian signature\footnote{For AdS/CFT in the Hamiltonian formalism for gravity and scalars and the role of the canonical momenta in correlation functions, see \cite{YK}.}. We define canonical 
momenta:
\be
\Pi^i={1\over\sqrt{-G}}{\d S\over\pa_rA_i}=-{1\over g^2}\,g^{rr}g^{ij}(\pa_rA_i-\pa_iA_r)~.
\ee
It is easy to see that, in the Fefferman-Graham coordinate system \ref{FG}, one gets the 
same Hamiltonian and action if one defines all quantities with respect to the conformally
rescaled metric $\ti G_{\m\n}={r^2\over\ell^2}\,G_{\m\n}$:
\be
E^i={1\over\sqrt{-g}}{\d S\over\d\pa_rA_i}=-{1\over g^2}\,g^{ij}(\pa_rA_j-\pa_jA_r)~.
\ee
This amounts to removing the factors of $g^{rr}$ from the formulas and raising and lowering
indices with the boundary metric $g_{ij}(r,x)$. This is the procedure we will adopt in what
follows, which of course is possible due to conformal invariance of the matter part 
of the action. From the analysis in Appendix \ref{holren}, equation \eq{FGexp}, it follows that
for a Ricci-flat boundary metric the first correction to the boundary metric appears at
order $r^3$ (this is the case, for instance, of the AdS black hole), therefore most boundary 
quantities we will be interested in are be unaffected by that.

The $\th$-term is a boundary term and does not contribute to the canonical momentum. The 
Hamiltonian now takes the following form:
\be
H=\int\dd^4x\left(-{g^2\over2}\,E^2-{1\over2g^2}\,F^2+E^i\pa_iA_r\right)-{\th\over8\pi^2}\int\dd^3x\,AF~.
\ee
Notice the unusual sign, which is due to the fact that we are doing radial quantization. The equations of motion give:
\bea
E_i&=&-{1\over g^2}\,(\pa_rA_i-\pa_iA_r)\nn
\pa_rE&=&{1\over g^2}\,*\dd F\nn
\na^iE_i&=&0~.
\eea
The action, written in first order form, now gives
\bea\label{1storderform}
S&=&\int\dd^4x\left(E^i(\pa_rA_i-\pa_iA_r)+{g^2\over2}\,E^2+{1\over2g^2}\,F^2-{\th\over4\pi^2}\,(\pa_rA)F\right)\nn
&=&\int\dd^4x\left(E^i(\pa_rA_i-\pa_iA_r)+{g^2\over2}\,E^2+{1\over2g^2}\,F^2\right)+{\th\over8\pi^2}\int\dd^3x\,AF~.
\eea
The bulk part of the action has the obvious symmetry $E'=F$, $F'=-E$, $g'=1/g$, $\th'=-g^4\th$. 
In the Euclidean signature action, the quadratic term in $E_i$ has a $-$ sign instead.

$A_r$ is a Lagrange multiplier. Its equation of motion gives the Gauss law
\be
\na^iE_i=0~.
\ee
We can solve the latter by
\be\label{top}
E^i(r,x)=\e^{ijk}\pa_jv_k(r,x)~
\ee
where $\e^{ijk}$ is the epsilon-tensor in the boundary metric $g_{ij}(r,x)$.
Notice that $E_i$ and $v_i$ are $r$-dependent. We will denote
\bea
v_i&=&v_i(x)=v_i(0,x)\nn
f_i&=&f_i(x)=E_i(0,x)~.
\eea
One can fix a gauge $A_r=0$, although this is not necessary.
As we mentioned earlier, there is a residual gauge invariance that leaves this gauge fixed, namely 
$A_\m\rightarrow A_\m+\pa_\m \vf$ where $\vf=\vf(x)$ has no $r$-dependence. These residual gauge transformations will act 
on the boundary value of the gauge field $A_i(x)$ and account for the gauge invariance of the dual CFT.

We fill in the equations of motion and get for the on-shell action:
\be\label{onshellwth}
S_{\sm{on-shell}}=\int\dd^3x\left(-\half\,EA+{\th\over8\pi^2}\,AF\right)~.
\ee

\subsection{Proof of $S$-duality}

We will give a proof of $S$-duality which, to our knowledge, is new. It relies on recasting the first order 
form of the action in a way that is manifestly $S$-duality invariant. We first remark that, since $E$ is a Lagrange multiplier, we can make the following redefinition:
\be\label{eredef}
{\sf E}=E-{\th\over4\pi^2}\,F~.
\ee
Since $F$ is conserved, this is compatible with a conservation constraint on $\sf E$. In fact, written as a function of $\sf E$ the action 
takes the form:
\be\label{1orderwcoupling}
S=\int\dd^4x\left({\sf E}^i(\pa_r A_i-\pa_iA_r)+\half\,g^2{\sf E}^2+{1\over2g^2}\left(1+{g^4\th^2\over(4\pi^2)^2}\right)\,F^2+{\th g^2\over4\pi^2}\,{\sf E}F\right)~.
\ee
With this rewriting, we have eliminated the boundary term. Indeed, the $\th$-angle in this form is not a boundary term.

As is easy to check, this form of the action has manifest $S$-duality invariance:
\bea\label{Sduality}
{\sf E}'&=&F\nn
F'&=&-{\sf E}\nn
\tau'&=&-{1\over\tau}~,
\eea
where
\bea\label{Scouplings}
\t&=&{\th\over4\pi^2}+{i\over g^2}=\tau_1+i\t_2\nn
{\th'\over4\pi^2}&=&-{{\th\over4\pi^2}\over {\th^2\over(4\pi^2)^2}+{1\over g^4}}=\t_1'=-{\t_2\over\t_1^2+\t_2^2}\nn
(g')^2&=&{1\over g^2}+{g^2\th^2\over(4\pi^2)^2}~~,~~\t_2'={\t_2\over\t_1^2+\t_2^2}~.
\eea
More precisely, we get
\be\label{trafowtheta}
S[A',{\sf E}']=S[A,{\sf E}]+\int\dd^3x\,{\sf E}A~.
\ee
In other words, in terms of $\sf E$ the boundary term that we get from the variation is the {\it usual} one that we get in the absence of 
the $\th$-angle.

One can see $S$-duality directly in the original first form order of the action \eq{1storderform}. We rewrite the $S$-duality transformations \eq{Sduality} in terms of the original variables 
\bea\label{dualEF}
E'&=&\left(1+{\th\th'\over(4\pi^2)^2}\right) F-{\th'\over4\pi^2}\,E\nn
F'&=&-E +{\th\over4\pi^2}\,F
\eea
where $\th'$ is the transformed coupling as in \eq{Scouplings}. Of course, this form of $S$-duality still satisfies $S^2=-1$, as it should. It is now easy to check that this leaves the action unchanged, up to boundary terms:
\be\label{SAE}
S[A',E']=S[A,E]+\int\dd^3x\left(E-{\th\over4\pi^2}\, F\right)A.
\ee
Of course, the extra terms are just the ones in \eq{trafowtheta}, as it should. This gives a direct proof of $S$-duality that doesn't involve any field redefinitions or integrating out fields.

It is interesting to note that the above can be simply rewritten as:
\be\label{SSA}
S[A',E']=S[A,E]-\int A'\wedge\dd A~.
\ee
This is in agreement with the definition of $S$-duality in \cite{Wittensl2z} from the CFT point of view. In fact, the above is like a Legendre transform between $A$ and a field of the same dimension $A'$.
Using the fact that $S$ squares to $-1$, we get
\be
S[A'']=S[A]~,
\ee
so we see that, whereas $S$ is not an invariance of the boundary theory, $S^2$ is. The above agrees with Witten's definition of $S$-duality in CFT \cite{Wittensl2z}.

\subsection{$SL(2,{\mathbb{R}})$ duality}\label{sl2rd}

We can generalize $SL(2,{\mathbb{Z}})$ invariance of the action to the full  $SL(2,{\mathbb{R}})$ that is relevant classically and plays a role in M-theory compactifications. Thus, we will generalize $S$-duality to include a full $SO(2)$ transformation. Let us define $g^*=g^*(g,\th)$ and $\th^*=\th^*(\th,g)$ as the couplings that we get after the $SL(2,\mathbb{R})$ transformation. We will take $E$ and $F$ to transform as
\bea
E^*&=&a\,E+b\,F\nn
F^*&=&c\,E+d\,F~.
\eea
Requiring invariance of the potential terms in \eq{1storderform} gives
\bea\label{SL2R}
g^*E^*&=&\eta\,g\,E+\d\,\sqrt{1-\eta^2}\,F/g\nn
F^*/g^*&=&-\d\,\sqrt{1-\eta^2}\,g\,E+\eta\,F/g
\eea
where $\d^2=1$ is an arbitrary sign which simply corresponds to the choice of sign in the square root. There is another sign that we have fixed here. The sign in front of $F$ on the r.h.s. of the above two equations is not fixed by invariance of the potential terms. It is fixed by invariance of the kinetic term (up to boundary terms), and the above choice is the one that renders the full bulk action invariant. $\eta$ is arbitrary and parametrizes the rotation part of the transformation. The transformed action is
\bea\label{actiontransform}
S[A^*,E^*]&=&S[A,E]+\int\dd^3x\left[\left({\d\over2}g^2\eta\sqrt{1-\eta^2}+g^2g^*{}^2{\th^*\over8\pi^2}(1-\eta^2)\right)fv\right.\nn
&+&\left.\left((1-\eta^2)-g^*{}^2{\th^*\over4\pi^2}\d\eta\sqrt{1-\eta^2}\right)fA+\left(-{\d\eta\over2g^2}\sqrt{1-\eta^2}+{g^*{}^2\over g^2}{\th^*\over8\pi^2}\eta^2-{\th\over8\pi^2}\right)FA\right]~.\nn
\eea

It is easy to check that the general transformation \eq{SL2R} automatically preserves the stress-energy tensor, as it should. 
It is also easy to see that the transformation \eq{SL2R} has unit determinant. 
In fact, it is most convenient to parametrize the above $SO(2)$ transformation as
\bea
\left(
\begin{array}{c}
g^*\,E^*  \\
F^*/g^* \end{array}
\right)&=&\left(
\begin{array}{cc}
\cos\a & \,\,\,\sin\a \\
-\sin\a & \,\,\,\cos\a
\end{array}
\right)
\left(
\begin{array}{c}
g\,E \\
F/g \end{array}
\right)~,
\eea
where without loss of generality we absorbed $\d$ in the sign of $\a$ and $\eta=\cos\a$. This transformation has an 
analytical continuation to Euclidean signature:
\bea\label{SL2RE}
\left(
\begin{array}{c}
g^*\,E^*  \\
F^*/g^* \end{array}
\right)&=&\left(
\begin{array}{cc}
\cosh\s & \,\,\,\sinh\s \\
\sinh\s & \,\,\,\cosh\s
\end{array}
\right)
\left(
\begin{array}{c}
g\,E \\
F/g \end{array}
\right)~,
\eea
which is a symmetry of the Euclidean bulk action up to boundary terms. The Euclidean transformation again preserves the stress tensor and $\dd E\wedge \dd F$. 

We also notice the regularity condition is invariant under the duality transformations.  
This can be seen as \eq{gaugeinvreg} implies
\be\label{regv}
F=-t\sqrt{-\Box}\Pi\,v~.
\ee
Under the transformation 
\be\label{regflow}
E^*=cE+sF=-c\sqrt{-\Box}\Pi\,A-t\,s{\sqrt{-\Box}}\Pi\,v=-\sqrt{-\Box}\Pi\,A^*
\ee
where implicit factors of $g,g^*$ are understood.

If we take $g^*=g'$ and $\th^*=\th'$ as in \eq{Scouplings}, we get back the $S$-duality transformation \eq{dualEF} if we set in addition
\bea\label{EFtrafo}
\cos\a={{g^2\th\over(4\pi)^2}\over\sqrt{1+{g^4\th^2\over(4\pi^2)^2}}}~,\quad \sin\a={1\over \sqrt{1+{g^4\th^2\over(4\pi^2)^2}}}\ .
\eea
Fill this in \eq{actiontransform} we recover \eq{SAE} as we should.
If we set instead $g^*=1/g$ and $\th'=-g^4\th$ then 
\be\label{flowend}
\sin\a=1
\ee
in order to obtain the S-dual transformation of \eq{1storderform}. Notice in both cases $g^*{}^2\th^*=-g^2\th$, which implements $S^2=-1$.

We note here that invariance of the bulk part of the action under the transformation \eq{SL2R} does not require the couplings to transform as \eq{Scouplings}. As we see, $g$ can actually be rescaled away, and $\th$ does not even enter \eq{SL2R}. To find out the transformation of $g$, the rest of the fields in the supermultiplet must be taken into account. In our case, as we will see in section \ref{ssusy} the relevant theory is the ${\cal N}=2$ supergravity in four dimensions, where $g\sim G_{\sm{N}}/\ell^2\sim 1/N^{3/2}$. Therefore, electric-magnetic duality will only be a symmetry of the action (up to boundary terms) if $g^*=g$. On the other hand, \eq{Scouplings} relates theories in asymptotically AdS$_4\times S^7$ of different size. In such a situation, the duality of \cite{BD1,BD2,Wittensl2z} which we prove here for any $g^*(g,\th),\th^*(g,\th)$ might actually be used to explore small AdS spaces.

Since $g^*(g,\th)$ is in general an arbitrary function, we can in fact get the $T$-generator from \eq{SL2R} by choosing $g^*=g$, $\cos\a=1$ and $\th^*=\th+1$. Using \eq{actiontransform}, this indeed simply shifts the $\theta$-term in the action. In this sense, our transformations are not $SO(2)$ but we in fact generate the full $SL(2,{\mathbb{R}})$.


\section{Generalized boundary conditions}\label{gbc}

Our goal is to study massive deformations of three-dimensional CFT's and how to describe them from the bulk. We will now develop the formalism to impose general boundary conditions for gauge fields and compute their on-shell action.

\subsection{Variation and consistent boundary conditions}

In the first order formalism we vary the action with respect to $E$ and $A$. The usual variational principle imposes on us a choice of Dirichlet or Neumann boundary conditions:
\be\label{bdyterm}
\d_AS_{\sm{bulk}}[A,E]=-\int\dd^3x\, E^i\,\d A_i|_{r=0} +({\mbox{eom}})~.
\ee
so that we get for the two independent fields:
\bea\label{DN}
\d A_i|_{r=0}&=&0~~~~~~{\mbox{Dirichlet}}\nn
E^i|_{r=0}&=&0~~~~~~{\mbox{Neumann}}~.
\eea
Since we are assuming that the boundary topology is trivial and we integrate out $A_r$, $E_i$ is expressed in terms of $v_i$ as in \eq{top} and there is no difference whether we vary with respect to $E$ or $v$.

The variation of the action \eq{bdyterm} also contains a contribution from the same boundary term evaluated at $r=\infty$. For regular solutions \eq{regulargauge}, using \eq{regularityr} this contribution is seen to be zero.

As explained in section \ref{scalardualities} for the scalar field case, in order to enforce general boundary conditions
we can add a boundary term that will modify the stationarity conditions \eq{DN}:
\be\label{newaction}
S[E,A]=S_{\sm{bulk}}[E,A]+S_{\sm{bdy}}[E,A]~.
\ee
The Dirichlet condition $\d A_i=0$ is obviously always a possible boundary condition. We will consider the generalization of the Neumann boundary condition. Varying \eq{newaction} and demanding that it be stationary, we find
\bea
\d_ES_{\sm{bdy}}[E,A]&=&0\nn
\d_AS_{\sm{bdy}}[E,A]&=&E^i(0,x)~.
\eea
In other words, in order for the action \eq{newaction} to be stationary under variations we need to require
\bea\label{bc}
{\d S_{\sm{bdy}}\over\d f_i}[f,A]&=&0\nn
{\d S_{\sm{bdy}}\over\d A_i}[f,A]&=&f^i~.
\eea
These two equations define our set of boundary conditions.

We can now obtain the on-shell effective action by adding the boundary term to \eq{onshellwth}:
\be\label{onshell}
S_{\sm{on-shell}}[f,A]=-\half\int\dd^3x\,A_if^i+S_{\sm{bdy}}[f,A]
~,
\ee
and $f$ and $A$ satisfy \eq{bc}.
As usual, we are assuming that there are no boundary contributions from $|x|=\infty$, while we keep boundary terms at $r=0$. As explained in section \ref{scalardualities} (see also section \ref{symf}), this can always be arranged by a choice of boundary terms or boundary conditions there \cite{don}. Unless the regularity condition at $r=\infty$ is imposed, the above only describes the contribution to the effective action coming from $r=0$. Notice that thanks to the classical conformal invariance of the bulk action, the on-shell action is completely finite at $r=0$ without the need to include any counterterms for the matter fields.

\subsection{General boundary action}\label{gba}




\subsubsection*{General boundary couplings}

We now construct the most general action with marginal operators and operators of dimension 4 that is gauge invariant, 
local, and covariant in $v$ and $A$\footnote{A priori there is no reason to require that the action be local in $v$ and 
$A$. In fact, the effective action will be non-local, as we discuss in section \ref{RGf}. The action 
described here corresponds to the simplest choice of local boundary conditions for the gauge field.}:
\be\label{generalaction}
S_{\sm{bdy}}[v,A]=\int\dd^3x\left(\half\a\, AF+\half\b\, vf+\g\,Af+\half\d\, F^2+\half\e\,f^2+\m\,fF -A\ti{\cal J}-fJ\right)
\ee
(for notation, see section \ref{notation}), where we leave the six coefficients arbitrary. $\ti{\cal J}$ is necessarily conserved by gauge invariance. This means there is a 
$\ti J$ such that $\ti{\cal J}=*\dd\ti J$. We will also define ${\cal J}=*\dd J$. 
The boundary conditions \eq{bc} derived from the above action take the form:
\bea\label{generaleom}
\b\,\Pi v+\g\,\Pi A+\e\,f+\m\,F&=&J\nn
\a\,F+(\g-1)\,f+\d\,*\dd F+\m\,*\dd f&=&\ti{\cal J}~,
\eea
both in Euclidean and in Lorentzian signature.

Taking the derivative of the first equation, we get 
\bea
\b\,f+\g\,F+\e\,*\dd f+\m\,*\dd F&=& {\cal J}\nn
\a\,F+(\g-1)\,f+\d\,*\dd F+\m\,*\dd f&=&\ti{\cal J}~.
\eea
The equations have an obvious symmetry under simultaneous exchange $f\leftrightarrow F$, $\a\leftrightarrow\b$, 
$\ve\leftrightarrow\d$ and $\gamma\leftrightarrow\g-1$. Notice that $\g$ plays a special role since it couples both 
equations. The two limiting cases $\g=0$ and $\g=1$ are in fact each other's $S$-duals. 

The above system can be reduced to
\bea\label{generaleom1}
a_1\,f+a_2\,F+a_3\,*\dd f&=&0\nn
b_1\,f+b_2\,F-a_3\,*\dd F&=&0~,
\eea
where the constants $a_i,b_i$ are given in \eq{a123} and we have set the sources to zero. It is possible to solve this system in general. One gets the following equation for $A$:
\be\label{generalmassiveeq}
(1+a\Box)\,A=b\,F~.
\ee
The values of $a,b$ are given in \eq{ab}. There is a similar equation for $f$, which for generic values of the constants $\a,\b,\g,\d,\e,\m$ is independent of the above. Again, for non-zero $b$ this is a (higher-derivative) massive equation for $A_i$. In fact, it is again the self-dual equation for the gauge field $A$, where the mass $m$ is now an operator, namely $m=(1+a\Box)/b\ve$.

In Appendix \ref{nearextremal} we give one further example, corresponding to near-extremal solutions.

\subsection{Symplectic flux through the boundary}\label{symf}

In section \ref{regularitysec} we derived the normalizability condition for Euclidean solutions. For non-tachyonic Lorentzian solutions, all solutions are normalizable, and the Euclidean boundary condition turns into an in-going boundary condition near the brane. In order to have a well-defined quantization problem, one also needs to ensure that the symplectic structure is finite and conserved \cite{marolf}. Whereas this is automatically ensured for Neumann or Dirichlet boundary conditions, it needs to be checked for generalized boundary conditions.

Given either of the boundary conditions
\bea
f_i&=&J_i^f\nn
A_i&=&J_i^A~,
\eea
vanishing of the symplectic flux in four dimensions requires that the matrices
\be\label{sym}
{\d J_i^f(x)\over\d A_j(x')}\quad\mbox{and}\quad{\d J_i^A(x)\over\d f_j(x')}
\ee
be symmetric. For our choice of boundary terms, the explicit expressions for $J_i^f$ and $J_i^A$ follow from \eq{generaleom}. It is then straightforward to check that the symmetry condition on \eq{sym} is met. Evaluating \eq{sym}, we encounter two types of structures: $\d_{ij}\d(x-x')$ and $(*\dd)_{ij}\d(x-x')$ (or the inverse of the latter). Both terms are symmetric. Symmetry of the latter is shown using partial integration.

In the scalar field case, there is an ultralocality requirement on the sources \eq{sym} \cite{marolf} in order to avoid partial derivatives contributing to the symplectic flux. For gauge fields, this requirement would seem to be spoiled by the presence of for example Chern-Simons terms. A modified symplectic structure can be defined \cite{Wittensl2z,marolf} such that there is no net contribution to the symplectic flux through the boundary.

\subsection{Supersymmetric boundary conditions}\label{ssusy}

Another interesting aspect of the generalized boundary conditions is which ones give solutions preserving certain amount of supersymmetry asymptotically. This then tells us which deformations of the boundary CFT are supersymmetric deformations. The conclusion however usually does depend on how we embed the truncated bosonic field content in a bigger supergravity theory. 

In our case the bosonic fields consist of only the graviton and abelian gauge field in asymptotically ${\rm{AdS}}_4$ space, in particular we have no scalars. The minimal theory to embed these as a consistent truncation is the the four dimensional gauged $\CN=2$ supergravity \cite{Freedman:1976aw} where our fields furnish the gravity multiplet. This is an extended supergravity theory with gauge symmetry $SO(2)=U(1)$ which can be further viewed as the consistent $SO(5)^V$ truncation of the gauged maximal $SO(8)$ supergravity \cite{Romans:1983qi}. This special $SO(5)$ invariance projects out all scalar fields and our theory is now indeed a sector of M-theory dimensionally reduced on ${\rm{S}}^7$. 

In order to find the supersymmetric boundary conditions, we follow the procedure of \cite{Hollands:2006zu}. First, since fermionic fields contribute to the symplectic flux through ${\rm AdS}$ boundary, we impose conditions on the gravitini fields so that the latter vanishes. Second, we find boundary conditions for the gauge fields consistent with supersymmetry transformations. We relegate the detailed analysis to Appendix \ref{susyapp} and simply present the conclusion here. We find that the only supersymmetric boundary condition for the gauge field is the Dirichlet boundary condition:
\be
A_i=0~,
\ee
which corresponds to the boundary action $S_{\sm{bdy}}=\int\dd^3x fA$, in Lorentzian signature.

\section{The flow of the boundary CFT}\label{fl}

In this section we will 
analyze the electric-magnetic flow of a general CFT with an AdS$_4$ dual with ${\cal N}=2$ supersymmetry. This is done by studying how generalized boundary conditions transform under electric-magnetic transformations. For convenience we will work in the $SL(2,{\mathbb{R}})$ formulation ($SL(1,1)$ in the Euclidean) relevant at the level of supergravity. Of course, quantum mechanically only a discrete subgroup of this is a symmetry. The latter is obtained by evaluating our formulas at discrete points.

\subsection{Flow of Dirichlet-Neumann boundary conditions}\label{clem}

We now investigate the properties of the effective action under $SL(2,\mathbb{R})$ transformations. Since $T$-transformations simply act as translations of $\th$, in this section we will work with the compact part, i.~e.~electric-magnetic rotations. For this purpose, we may take $\th^*=\th=0$. In the Lorentzian case, electric-magnetic duality acts as $SO(2)$ transformations. As seen above, the relevant pair is not $(E,A)$, but $(v,A)$ and their corresponding dual field strengths $(E,F)$. In the Euclidean case, we get $SO(1,1)$ instead. 

The relevant transformation is
\bea\label{lorenflow}
g^*v_i^*&=&c\, gv_i+s\,{A_i\over g}\nn
{A_i^*\over g^*}&=&ts\, gv_i+c\,{A_i\over g}~.
\eea
Here we recall $t=-1$ for Lorentzian signature and $1$ for Euclidean. Also $c,s$ stands for $\cos\a,\sin\a$ in the former and $\cosh\s,\sinh\s$ in the latter. Taking the curl of this rotates electric and magnetic fields into each other. In the Euclidean case these are boosts, so we will loosely refer to them as the $SO(1,1)$ ``flow'' from now on. The first order bulk action is invariant under this up to a boundary term \eq{actiontransform}, which present below:
\be\label{actionemrot}
S_{\sm{bulk}}[v^*,A^*]-S_{\sm{bulk}}[v,A]=-t\int\left(s^2\,A\wedge\dd v+\half sc(g^2\,v\wedge\dd v+{t\over g^2}A\wedge\dd A)\right)~.
\ee
Of course, the boundary term $S_{\sm{bdy}}[v,A]$ of the effective action will transform as well, and its transformation properties need to be analyzed case by case. We will do this after introducing the general boundary conditions.

By minimizing the $SO(2)$-transformed action, it is not hard to check that the boundary conditions required for stationarity satisfy:
\be
A_i+{c\over s}\, v_i=0~,
\ee
This boundary condition is still linear, yet in contrast to \eq{dtbc} it preserves conformal invariance as it does not introduce a new scale. Thus, introducing the $SO(2)$ flow we probe a one-parameter family of conformally invariant boundary conditions. The $SO(2)$ flow corresponds to a 1-parameter deformation of boundary theories.

It is clear that for $c/s=\infty$ we get the Neumann boundary condition, and for $c/s=0$ the Dirichlet one.
We started with a situation where $A$ was integrated over but $v$ was fixed such that $F_{ij}[v]$ is a flat connection. After applying the $S$-dual transformation, we get a new boundary condition where $A$ is fixed whereas $v$ becomes dynamical. This precisely parallels the CFT discussion in \cite{Wittensl2z}.

There is a unique marginal purely three dimensional action that is invariant under the electric-magnetic duality \eq{lorenflow}. It is the difference (Euclidean) or sum (Lorentzian) of two Chern-Simons terms:
\be\label{invariant}
S_{\sm{invariant}}[v,A]=\int\dd^3 x\left(v\wedge\dd v-tA\wedge\dd A\right)~.
\ee

\subsection{$SL(2,\mathbb{R})$ transformed boundary conditions}

We will now include the $\th$-term and analyze the transformation of general boundary conditions.
The bulk action changes by (see \eq{actiontransform})
\bea\label{actionflow}
S[A^*,E^*]&=&S[A,E]+\int\dd^3x\left[\left(-{t\over2}g^2\,cs+g^2g^*{}^2{\th^*\over8\pi^2}\,s^2\right)fv\right.\nn
&-&\left.t\left(s^2-{g^*{}^2\th^*\over4\pi^2}\,cs\right)fA+\left(-{cs\over2g^2}+{g^*{}^2\over g^2}{\th^*\over8\pi^2}\,c^2-{\th\over8\pi^2}\right)FA\right]~.\nn
\eea
The corresponding equations of variation are
\bea\label{slflow}
{1\over g}\,{\d S^*_{\sm{bdy}}\over \d f_i}&=&ts\left(g^*v^*-{g^*{}^2\th^*\over4\pi^2}{A^*\over g^*}\right)\,\nn
g\,{\d S^*_{\sm{bdy}}\over\d A_i}&=& c\left(g^*f^*-{g^*{}^2\th^*\over4\pi^2}{F^*\over g^*}\right)~.
\eea
This can be re-organized in both signatures to
\bea\label{slcomu}
{\d S^*_{\sm{bdy}}\over \d f_i^*}&=&0\nn
{\d S^*_{\sm{bdy}}\over\d A^*_i}&=&f^*-{\th^*\over4\pi^2}F^*=\sf f^*~.
\eea
Thanks to the fact that we are working off-shell, \eq{slcomu} says that extremization commutes with duality transformations. 

We can further simplify our discussion by viewing the $\th$-term as a term in the boundary action. We may now keep factors of $g$ implicit and use \eq{actionemrot} to find  
\bea\label{floweqa}
{\d S^*_{\sm{bdy}}\over \d f_i}&=&ts\left(c\,v+s\,{A}\right)\,\nn
{\d S^*_{\sm{bdy}}\over\d A_i}&=& c\left(c\,f+s\,{F}\right)~.
\eea
Since these transformations interpolate between different boundary effective actions, we will refer to this as the flow of the boundary theory (which is not to be confused with RG flow, although in the next section we will explain how they are connected). 
 The flow starts at $\s=0$, $s=0$, $c=1$, where it of course reduces to \eq{bc}. The Lorentzian flow swaps the value of $c,s$ at the end. In contrast, the Euclidean flow ends at $\sigma=-\infty$ where $s=-c$ blows up. 
 
\subsection{Massive deformation}

We now come to the case of massive deformations, given by  the linear boundary condition \eq{dtbc} which explicitly breaks conformal invariance. In fact, it adds a relevant operator to the action. It corresponds to the following choice of boundary term:
\be\label{relevantdef}
S_{\sm{bdy}}[f,A]=\int\dd^3x\left(A_i f_i-{1\over2m}\, f_i^2\right)~.
\ee
Without the deformation term, this action simply imposes the Dirichlet boundary condition $A=0$. It is easy to check that after the flow we get again a single independent equation
\be
c\,A_i+ts\,v={1\over m}\left(c\,f_i+s\,F_i\right)~,
\ee
which is of course nothing else than the flow of the linear boundary condition \eq{dtbc}.
In the Euclidean case, it is interesting to note that in the limit $\s\rightarrow\infty$ the above reduces to
\be
\hat A_i=-{1\over m}\,\e_{ijk}\pa_j\hat A_k
\ee
where $\hat A_i=A_i-v_i$. Hence, in the limit $\s\rightarrow\infty$ we get the self-dual massive theory of \cite{PTvN}!

In Lorentzian signature, we can continue the flow to $c=0,s=1$ where the boundary condition becomes 
\be\label{srelevant}
v=-{1\over m}\,F~,
\ee
which is the linear boundary condition on the $S$-dual fields.

\subsection{Self-dual boundary conditions}

We will now study the self-dual case, which corresponds to the equations of motion of the topologically massive theory \cite{PTvN}. We look for a choice of boundary terms that combines the self-dual and massive boundary conditions leading to \eq{SD}. This is achieved by the choice:
\be\label{fF}
S_{\sm{bdy}}[A,f]=-\int\dd^3x\left({g^2\over2m}\,f^2+{\ve\over m}\,fF\right)~,
\ee
where we have reinstated factors of $g$ left out in \eq{sdbc}, and for simplicity we first consider $\th=0$. After some algebra, we find that the modified equations of motion combined together lead to
\bea\label{sdtransform}
f&=&-{\ve\over g^2}\,F\nn
A&=&{\ve\over m}\,F~,
\eea
that is, precisely the equations of motion of the ``master action'' for the self-dual theory! \cite{DJ}. It is remarkable that all $\s$-dependence in \eq{sdtransform} has disappeared. Another way to see this is to simply let the flow act directly on \eq{sdtransform}. It is easy to see these equations are invariant under it. So we find that the self-dual theory remains invariant during the whole flow! This is exactly what we expect: if we start with the self-dual boundary condition \eq{sdbc} which is a solution of the bulk self-duality equation \eq{SD}, the solution has vanishing bulk stress-energy tensor:
\be\label{tmn=0}
T_{\m\n}=0~.
\ee
Now since electric-magnetic duality preserves the form of the stress-energy tensor, \eq{tmn=0} should be true throughout the flow. This is exactly what we find: \eq{sdtransform} are valid throughout, and this ensures vanishing stress tensor. 
Of course, the above action again contains relevant deformations since the theory is massive. 

We should notice here that, although the equations of motion \eq{sdtransform} are invariant under $SO(1,1)$, the action \eq{fF} (or, for that matter, the full action including the bulk term) is not invariant under it. In Appendix \ref{sdactions} we give two other choices of boundary terms that give the same boundary conditions. These choices should be continuously related to the one in \eq{fF} by the flow.

This discussion should shed more light on the ``self-duality'' property of this three-dimensional theory: this is the boundary theory that corresponds to a choice of boundary conditions that leaves the action invariant under four-dimensional electric-magnetic transformations. Notice that the bulk part of the action is essential in checking this invariance. It is also easy to check that none of the purely three-dimensional actions \cite{DJ} are invariant under \eq{lorenflow}. They are only invariant under the transformation in \cite{hagen}, which is different from electric-magnetic duality in the bulk. In fact, the unique boundary invariant at this order was written in equation \eq{invariant}, which is simply two non-interacting copies of abelian Chern-Simons theory. The point is that the effective action receives a boundary contribution coming from the variation of the bulk term $\int\dd^4x\,E_i\pa_r A_i$.

\subsection{General case}\label{generalflow}

We are now ready to study the more general boundary action \eq{generalaction}. For simplicity we will set $g=1$. The transformed equations \eq{generaleom} read
\bea\label{generalflow}
\b\,v'+\g\,A'+\e\,f'+\m\,F'&=&0\nn
\a\,F'+(\g-1)\,f'+\d\,*\dd F'+\m\,*\dd f'&=&0~.
\eea
Performing linear combinations as in \eq{generaleom1}
leads to 
\bea\label{generalflow2}
a_1'\,f+a_2'\,F+a_3'\,*\dd f&=&0\nn
b_1'\,f+b_2'\,F-a_3'\,*\dd F&=&0~,
\eea
where we have the transformed coefficients
\bea\label{flowcoeff}
&&a_1'=a_1c^2+t(a_2+tb_1)cs+tb_2s^2\nn
&&a_2'=a_2c^2+(a_1+b_2)cs+b_1s^2\nn
&&b_1'=b_1c^2+t(a_1+b_2)cs+a_2s^2\nn
&&b_2'=b_2c^2+t(a_2+tb_1)cs+ta_1s^2\nn
&&a_3'=a_3~.
\eea

Remarkably it can be shown that the following are invariants under the flow \eq{flowcoeff}
\bea\label{invariants}
&& a_1'-b_2'=a_1-b_2,\quad a_2'-t\,b_1'=a_2-t\,b_1,\quad a_3'=a_3,\nn
&& a_1'b_2'-a_2'b_1'=a_1b_2-a_2b_1,\quad c_1'd_2'-c_2'd_1'=0~.
\eea
Using the definition of $a,b$  in \eq{ab} we see that these are invariants as well. This means under the flow, the equation for $A$ remains unchanged
\be
(1+a\Box)A=bF~.
\ee
From this, we conclude that for generic values of coupling constants, the equations \eq{generaleom} and hence the boundary conditions are invariant under the flow! We should stress though that the effective action \eq{generalaction} itself is in general not invariant and hence the invariance of the equation is not a consequence of
the invariance of the action.This is reminiscent of the realization of EM
duality in the 4d theory!

This would seem to contradict what we found above where there were nontrivial flows of 
boundary condition. However recall $a,b$ are well-defined only if $a_1b_2-a_2b_1\ne0$. In 
fact, it's easy to check that actions given in \eq{relevantdef} \eq{fF} 
\eq{moresdac}, as well as the trivial case $S_{\sm{bdy}}=0$, all have $a_1b_2-a_2b_1=0$. 


\section{RG-flows, $S$-duality, and particle-vortex duality}\label{RGf}

Now we come to the interplay between $S$-duality and RG-flows. We will first compute the one- and two-point functions using the method of holographic renormalization, and then explain how $S$-duality acts on them. We will then discuss the physical interpretation, which requires considering both electric and magnetic sources in the AdS/CFT dictionary. Finally we will discuss massive deformations leading to RG-flows from one theory to its $S$-dual, and compute the effective action.

\subsection{$S$-duality of the two-point function and Legendre transforms}

\subsubsection*{Legendre transformed one-point function}

We can now easily compute the 1- and 2-point functions. We use the method of holographic renormalization \cite{dHSS} (see also \cite{Skenderis}). After solving the bulk equations of motion, one regularizes the action and adds counterterms. To compute the 1-point function, one needs to take the derivative of the action with respect to the source $A_{i(0)}(x)$. Doing this in general is non-trivial, as the dependence of $f_i(x)$ on $A_{i(0)}$ depends on the particular solution. The way to do this in general is to write the boundary action in terms of the regulated field $A_i(\e,x)$. We find for the dimension-2 current:
\be\label{dim2}
\d S=\int\dd^3x\left[{1\over g^2}\,(\pa_rA_i(r,x)-\pa_iA_r(r,x))+{\th\over4\pi^2}\,F_i(r,x)\right]\d A_i(r,x)|_{r=\e}~,
\ee
where we have used the equations of motion. The above variation is completely finite, as it should. It is shown in section \ref{backreaction} how to renormalize the action, including the coupling to gravity\footnote{In usual holographic renormalization, one should also use the induced boundary metric $\g_{ij}(\e,x)=1/\e^2g_{ij}(\e,x)$. Thanks to conformal invariance of the matter part of the action the powers of $\e$ drop out. See section \ref{1of}}. We are left with:
\bea\label{1pt}
\bra {\cal O}_{2i}(x)\ket_A&=&\lim_{\e\rightarrow0}{\d S[A]\over\d A_i(\e,x)}=\left(-E_i(r,x)+{\th\over4\pi^2}\,F_i(r,x)\right)|_{r=0}=-f_i(x)+{\th\over4\pi^2}\,F_i(x)\nn
&=&-{\sf f_i}(x)~.
\eea
The first term is the usual one, see \eq{recipe}. The second term is the contribution from the $\th$-term. The $\th$-term does not modify the equations of motion in the bulk. However, it does modify the on-shell action, and therefore it contributes to the expectation value of the conserved current.


In the theory where we have the operator of dimension 2, we have shown that the generating functional of connected correlation functions is: $W=-S_{\sm{on-shell}}=\half\int\dd^3x\,{\sf f} A$.
One can check that the effective action is minus this $\G=S_{\sm{on-shell}}$.

Now as we discussed in section \ref{scalardualities}, we can get to the dual CFT if we do a Legendre transform. The roles of the effective action and the generating functional are then interchanged. In particular, for a linear choice of boundary conditions we have
\bea
\ti\G&=&W\nn
\ti W&=&\G~,
\eea
where the right-hand side should be viewed as a function of the source $A$, and the left-hand side as a function of the dual source $\ti f$.
It is easy to check that the one-point function is independent of the linear choice of boundary condition. That 
is basically what the generic holographic result \eq{dim2} tells us. For any such choice we get:
\be\label{1ptt}
\bra \ti {\cal O}_1\ket_{\sf f}=-{\d\ti W[\sf f]\over\d\sf f}=-\Pi\,A
\ee
as we would expect. This is a gauge field of dimension 1, projected onto its transverse part. From this we now get
the conserved current of the dual theory, 
\be
\bra\ti{\cal O}_2\ket_{\sf f}=-F~.
\ee

\subsubsection*{$S$-duality of the one- and two-point functions}

We saw that the action $S[A,E]$ and its $S$-dual $S[A',E']$ are each other's Legendre transforms, \eq{SSA}. This explains the minus sign in comparing $S[A]$ \eq{Aaction} with its $S$-dual \eq{Sd}. The generating functional of connected correlation functions of the $S$-dual theory thus coincides with the Legendre transformed functional, with the $S$-dual couplings:
\be\label{Wp}
W'[{\sf f}']=\ti W[{\sf f}';g',\th']~.
\ee
Let us spell this out in more detail in this case. Recall the definition of the generator of connected correlation functions: $-W[A]=S_{\sm{on-shell}}[A]=-\half\int{\sf f}A$, where ${\sf f}={\sf f}[A]$. Define now the Legendre transform:
\be\label{lW}
\ti W[\ti A]=W[A]+\int\ti A F~,
\ee
therefore
\be\label{tiW}
{\d\ti W\over\d\ti A}=F~~,~~{\d W\over\d A}=-\ti F~.
\ee
Using \eq{tiW} and filling the value of $W[A]$ in \eq{lW}, we find in this case $\ti W[\ti A]=-\half\int{\sf f}A=S[A]$. From \eq{tiW} we then precisely find the definition of $S$-duality \eq{Sduality}, as it should. This justifies \eq{Wp} and agrees with what we found in \eq{SSA}. We rewrite the dual potential as:
\be
\ti W[\ti A]=\half\int\ti{\sf f}\ti A=-\ti S[\ti A]~.
\ee
This is just what we expect -- the dual CFT is related to the dual bulk action in the standard way.

The one-point function of the $S$-dual conserved current is
\be\label{1pfd}
\bra {\cal O}_{2i}'\ket_{A'}={\d S'\over\d A'}=-F=-{\sf f}'~.
\ee

We now compute the two-point function and its $S$-dual. The two-point function is defined as

\bea
\bra {\cal O}_{2i}(x){\cal O}_{2j}(x')\ket_{A=0}&=&{\d\over\d A_i(x)}{\d\over\d A_j(x')}\,S_{\sm{on-shell}}|_{A=0}\nn
&=&{\sqrt{-\Box}\over g^2}\,\Pi_{ij}(x-x')+{\th\over4\pi^2}\,(*\dd)_{ij}\d(x-x')~.
\eea
In momentum space,
\be\label{2ptem}
\bra {\cal O}_{2i}(p){\cal O}_{2i}(-p)\ket_{A=0}={1\over g^2}|p|\Pi_{ij}-{\th\over(4\pi^2)^2}i\e_{ijk}p_k~.
\ee
Since the right-hand side only depends on the couplings, in order to compute the $S$-dual of this we might just apply \eq{Scouplings} and 
get the $S$-dual result in \cite{Wittensl2z,tassos}. However, our formalism allows us to perform a check from first principles, which is what
we will now do. This will be useful both in order to understand the structure of $S$-duality, and apply it in the general case.

The definition of the two-point function in the $S$-dual theory is simply:
\be
\bra{\cal O}_{2i}'{\cal O}_{2j}'\ket={\d\over\d A_i'}{\d\over\d A'_j}\,S'_{\sm{on-shell}}~.
\ee
We use the transformations \eq{dualEF} to compute the $S$-dual of the on-shell action
\bea
S'_{\sm{on-shell}}&=&-\half\int\dd^3x\,{\sf E}' A'=\half\int\dd^3x\,EA-{\th\over4\pi^2}\,AF\nn
&=&\half\int\dd^3x\left(-{1\over g^2}\,AM\Pi A-{\th\over4\pi^2}\,A*\dd A\right)~,
\eea
where in the last line we have used the in-going boundary condition. This is, as announced, minus the on-shell action.

We now solve \eq{dualEF} and write the functional derivative as follows:
\be
{\d\over\d A_i'}=-\left({\th'\over4\pi^2}+g^2\left(1+{\th\th'\over(4\pi^2)^2}\right)M^{-1}*\dd\right)_{ij}{\d\over\d A_j}~,
\ee
where we have used the first of the identities \eq{id4}.

Since $\Pi$ is non-invertible, these identities should be understood as follows. One adds a gauge-fixing term to the action and inverting the
resulting modified propagator. The final result does not contain any inverse propagators, and one can safely take the limit that the
gauge-fixing terms go to zero. We get:
\bea\label{dual}
\bra {\cal O}_{2i}'{\cal O}_{2j}'\ket &=&{\d\over\d A_i'}{\d\over\d A_j'}\,S'\nn
&=&\left(-{\th'\over4\pi^2}+g^2\left(1+{\th\th'\over(4\pi^2)^2}\right)M^{-1}*\dd\right)_{ik}\left(-{\th'\over4\pi^2}
+g^2\left(1+{\th\th'\over(4\pi^2)^2}\right)M^{-1}*\dd\right)_{jl}\nn
&\times&{\d\over\d A_k}{\d\over\d A_l}S
\nn
&=&\left({\th'^2\over(4\pi^2)^2 g^2}+g^2\left(1+{\th\th'\over(4\pi^2)^2}\right)^2\right)M\Pi-\left({\th\th'^2\over(4\pi^2)^3}+{\th\over4\pi^2} g^4(1+{\th\th'\over(4\pi^2)^2})^2\right)*\dd\nn
&=&{g^2\over1+g^2{\th^2\over(4\pi^2)^2}}\,M\Pi-{{g^4\th\over4\pi^2}\over1+{g^4\th^2\over(4\pi^2)^2}}\,*\dd
\eea
which is indeed the $S$-dual transform in the usual sense.

\subsection{Physical interpretation of S-duality}\label{phys}

We have seen how $S$-duality acts in the bulk and on the boundary, and we now come to the physical interpretation. 

In this section we will deal with the pure Dirichlet/Neumann problems, therefore it will not be necessary to add any additional boundary terms. More general cases will be worked out in the next section.

We saw in section \ref{gba} that the parameter $\g$ interpolates between mutually $S$-dual solutions. This is easy to see from \eq{generaleom}: the case $\g=1$ corresponds to the Dirichlet boundary condition, whereas $\g=0$ is the Neumann boundary condition. Inclusion of the sources $J$ and $\ti{\cal J}$ will clarify what this $S$-duality actually means for the boundary CFT. Indeed, \eq{generaleom} is not quite symmetric with respect to $J$ and $\ti{\cal J}$: $J$ has dimension 1, whereas $\ti {\cal J}$ has dimension 2. Therefore, there exists a $\ti J$ such that $\ti {\cal J}=*\dd\ti J$. We can then rewrite the source terms in the following symmetric form:
\be\label{ems}
S_{\sm{source}}=-\int\dd^3x\left(F\ti J+fJ\right)=-\int\dd^3x\left(A\ti{\cal J}+fJ\right)~.
\ee
Although $\ti J$ is not necessarily conserved, this expression is gauge invariant and hence well-defined. In 
a theory with gauge fields, this expression has a natural interpretation. Let us promote $J$ and $\ti J$ to bulk
sources $J(r,x)$, $\ti J(r,x)$ localized at the boundary, with a delta-function at $r=0$. ${\cal J}$ is 
then the magnetic source that couples to the bulk electric field $E(r,x)$. $\ti J(r,x)$, on the other hand,
acts as a source for the magnetic field in the directions transverse to it. This is easy to see: if the sources would contain any bulk contributions, they would modify the equations of motion as
\bea
E&=&-{1\over g^2}\,\pa_rA+{1\over g^2}\,J\nn
\pa_rE&=&{1\over g^2}\,*\dd F-\ti J~.
\eea
Clearly then, $\ti J$ will contribute to the curl of the magnetic field, and $*\dd J$ will contribute to the curl of the electric field. We write $(J_e,J_m)=(\ti{\cal J},{\cal J})$. In this paper we will not consider any bulk contributions coming from the sources. When restricted to the boundary, both sources have dimension 2. However, since they are
conserved we can write $(\ti{\cal J},{\cal J})=*\dd(\ti J,J)$ where $(\ti J,J)$ have dimension 1 in the Legendre transformed theory.
Bulk electric-magnetic duality interchanges them: $(\ti J',J')=(J,-\ti J)$.

In AdS/CFT with the usual boundary conditions, $J$ is interpreted as the source in the boundary CFT.
However, the above makes clear that, in the presence of both electric and magnetic charges, the theory
can also be deformed by $\ti J$, hence it ought to have a boundary interpretation. Since it has dimension
1, this suggests that it is dual to an operator of dimension 2, and in fact this operator should be 
$S$-dual to ${\cal O}_2$. Furthermore, from the existence of an alternative quantization scheme where
the boundary operator has dimension 1, one would guess that the bulk sources $\ti{\cal J}$ and ${\cal J}$
themselves (restricted to the boundary) have an interpretation in terms of dual operators of dimension 1.
This is in fact very natural, as ${\cal J}$ and $\ti{\cal J}$ are conserved so they might give rise to
a boundary operator which is the gauge field itself. We will now verify that this is indeed the case. 

\begin{figure}[t]
\begin{center}
\includegraphics[height=4in,width=5in]{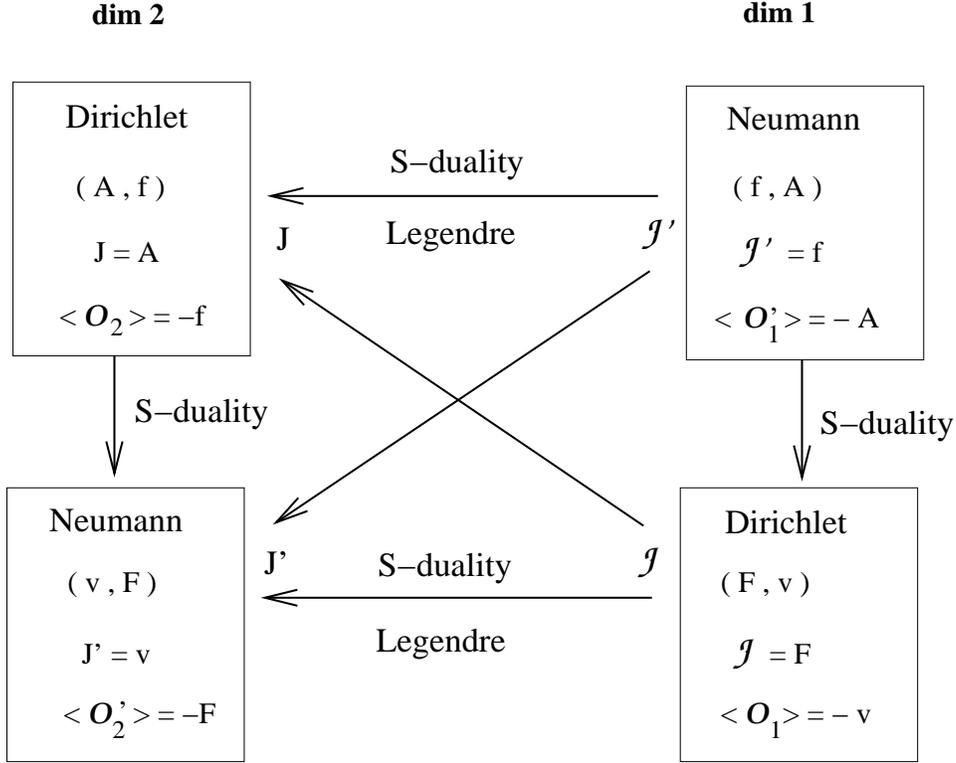}
\caption{\small Diagram of the various S-duality transformations.}
\label{diagram}
\end{center}
\end{figure}

We start with the on-shell action $S_{\sm{on-shell}}=-\half\int\dd^3x\,{\sf f}A$. Now instead of fixing
the gauge field $A$, we will fix the electric field by a source of dimension 1:
\be
\ti J=-J'=-A'={\sf v}~.
\ee
The prime suggests that this is the $S$-dual to $J$, however at this stage we are not making use of 
$S$-duality and we regard the above purely as a boundary problem for ${\sf{v}}$. Clearly, the dual operator will have dimension 2. We compute:
\be
\bra{\cal O}_2'\ket=-{\d W'\over\d{\sf v}}=-F~,
\ee
where we have used $W'=S'_{\sm{on-shell}}=\half\int\dd^3x\,{\sf f}A$. This result agrees with \eq{1pfd}. We now compute the two-point function. To do that, we use regularity to write $F=M\Pi v$ and take a further derivative with respect to ${\sf v}$. In order to do that, we invert ${\sf v}=v-\th A$. We get:
\be
v={1\over1+{\th^2\over(4\pi^2)^2}}\left({\sf v}-{\th\over4\pi^2 M}\,*\dd{\sf v}\right)~.
\ee
So we get
\be
\bra{\cal O}_2'{\cal O}_2'\ket={M\over1+{\th^2\over(4\pi^2)^2}}\,\Pi-{{\th\over4\pi^2}\over1+{\th^2\over(4\pi^2)^2}}\,*\dd~.
\ee
This is exactly the $S$-dual \eq{dual} of the two-point function of a conserved current! (we have set 
$g=1$). Notice that we have not assumed any knowledge of $S$-duality here: it simply follows from the fact that we identify $\ti J$ as the source. This describes the lower-left corner of figure \ref{diagram}.

The upper right corner of figure \ref{diagram} is the (equivalent) description in terms of the dual gauge field:
this is obtained by taking the dual source ${\cal J}'=-\ti{\cal J}$ and fixing:
\be
\ti{\cal J}={\sf f}.
\ee
We find:
\bea
\bra{\cal O}_1'\ket_{\cal J}&=&\Pi A\nn
\bra{\cal O}_1'{\cal O}'_1\ket&=&{1\over M\left(1+{\th^2\over(4\pi^2)^2}\right)}\,\Pi-{{\th\over4\pi^2}\over M^2\left(1+{\th^2\over(4\pi^2)^2}\right)}\,*\dd~.
\eea
This is indeed the $S$-dual of the two-point function of a gauge field and agrees with \eq{1ptt}.  Of course, taking the derivative of the above we 
recover the correlation functions of the dual current. This is another
check that the method of dual sources gives correct results.

We now complete the Dirichlet case on the main diagonal of \ref{diagram}. The upper-left corner is the usual Dirichlet boundary condition $A=J$. The lower-right case is obtained identifying $F={\cal J}$. The one- and two-point functions are then:
\bea
\bra{\cal O}_1\ket&=&-\Pi{\sf v}\nn
\bra{\cal O}_1{\cal O}_1\ket&=&{1\over M}\,\Pi-{\th\over4\pi^2 M^2}\,*\dd~.
\eea
There is no $S$-duality transform here, as it should. Once more, taking the derivative we get the usual current.

We can now rephrase physically the situation in Figure \ref{diagram}. First of all, notice that the Legendre 
transform and $S$-duality transformations that translate us vertically or horizontally are the same. In both cases, 
the term that we add to the action is $\int{\sf f}A$. Only in one case (the horizontal one) this is regarded as a 
coupling between the source and the current, whereas in the vertical case it is a Chern-Simons coupling between the 
background field and the new dynamical gauge field. Obviously, the diagonal action simply corresponds to the action 
of $*\dd$ and ensures that the theory is unitary. The difference between $S$-dual theories lies in what we 
interpret as a source and as an operator. In the upper-left corner, we fix as a source the bulk gauge field $A$. The 
corresponding CFT operator that couples to it is then an induced current of dimension 2. In the upper-right, on the 
other hand, $A$ has become a dynamical field with corresponding conserved current $F$, and the background field $v$ 
is fixed via its curl $f$. This is the particle-vortex duality. In the next subsection we will see how this duality 
is achieved by an RG-flow. From the bulk point of view, particle-vortex duality corresponds to interchanging $J$ and 
${\cal J}'$ as we have shown. The latter fact may have applications to other situations, such as the non-abelian case 
(see section 6.5).

\subsection{RG-flow of the two-point function and $S$-duality}


It is known from field theory that massive deformations of three-dimensional CFT's by double-trace operators lead to RG flows from the UV towards the IR fixed point. In \cite{tassos} it was shown using field theory arguments that if the IR dimension of a conserved current is 2, in the UV it corresponds to an operator that is $S$-dual to a gauge field of dimension 1. In this section we will show how this result comes about from the bulk point of view, and how it fits in the general picture of Figure \ref{diagram}.

We will modify the massive boundary condition to include a $\th$-angle:
\be
A+{1\over m}\,{\sf f}=J~,
\ee
in other words
\be\label{massive}
A-{\th\over m}\,F+{1\over m}\,f=J~.
\ee
As our second boundary condition, we take
\be
f=cM\Pi A~.
\ee
We will choose $c=-1$, which corresponds (for the tachyonic case) to the regular configurations, as explained in section \ref{regularitysec}. The value of $c$ only results in an overall rescaling of the two-point function, however $c$ has to be negative in order for the two-point function to be positive definite, as explained in section \ref{sdtmt}. We obtain the above boundary conditions if we choose $\g=1$, $\e=1/m$, $\m=-\th/m$, $\a=-M\th/m$ in \eq{generaleom}, and the other coefficients zero (in the rest of this section we rescale $\th$ by a factor of $4\pi^2$ compared to our previous formulas). The on-shell action is:
\be\label{osA}
S_{\sm{on-shell}}=\half\int\left(-{M\over m}(m-M)A\Pi A+{M\th\over m}\,A\wedge\dd A+2MJA\right)~.
\ee
Now we solve for $A$ in \eq{massive} in terms of the source. We get:
\bea
A&=&{m\over\G}\left((m-M)\,J+\th*\dd J\right)\nn
\G&=&(m-M)^2-t\th^2M^2~.
\eea
Plugging this back in the action, we get
\be\label{mons}
S_{\sm{on-shell}}=\int{mM\over2\G}\left(J(m-M)\Pi J+\th J\wedge\dd J\right)~.
\ee
The one-point function is easily computed and found to be $\bra{\cal O}_2\ket=MA=-f$. It is clear that the IR limit $|p|\ll m$ of the two-point function is the usual one for a conserved current of dimension 2, \eq{2ptem}. On the other hand, it is quite remarkable that in the UV limit we get
\be\label{2ptUV}
\bra{\cal O}_{2i}{\cal O}_{2j}\ket={m\over1+\th^2}\left(\Pi_{ij}-\th {i\e_{ijk}p_k\over|p|}\right)+{m^2\over|p|(1+\th^2)^2} \left((1-\th^2)\,\Pi_{ij}-2\th {i\e_{ijk}p_k\over|p|}\right)~.
\ee
The second term in \eq{2ptUV} was also found in \cite{tassos} and corresponds to the ($S$-dual) two-point 
function of an operator of dimension 1 -- a gauge field. This is what we expect. 
Notice that, once again, there is no assumption about $S$-duality made; the massive deformation automatically leads us to 
the upper-right part of the diagram \ref{diagram}.

The first term in \eq{2ptUV} is a contact term \cite{tassos}. Naively it would correspond to an operator with UV dimension 3/2. However, a spin one 
operator cannot have dimension 3/2 in a conformal theory, therefore it cannot be a chiral primary and it must be 
a descent operator. There is indeed such an operator in QED in 3 dimensions \cite{BKW,KS}. In that case, the 
current gets dimension 3/2 and it can be written as $f_i=\pa_i\f$ where $\f$ is a scalar field. However, the 
presence of such a term without its primary field in the OPE breaks conformal invariance in the UV.

The result can be given a different interpretation that agrees with what we said earlier if we take 
${\cal J}=*\dd J$ as the source. In that case, we find the two-point function of $\Pi{\sf{v}}$ with itself, and 
we are in the 
lower-right part of \ref{diagram}! In particular, the $S$-duality transform is absent here and will instead 
appear in the IR.

It is straightforward to check that the method to get the $S$-dual two-point functions outlined in section \ref{phys} works in this case as well. If we fix the boundary value of the gauge field, $J=A$, we easily find the two-point function from \eq{osA}. If instead we couple the theory to an electric source $\ti{\cal J}$  as in \eq{ems} and take derivatives of \eq{osA} with respect to $\ti{\cal J}$, we get the $S$-dual two-point function. This explains the $S$-duality in the flow when we turn on the mixed boundary condition \eq{massive}: in the IR limit we are fixing $A$ and $J$ is the usual magnetic source; in the UV, the second term dominates and $J$ becomes an electric source! Thus, \eq{massive} interpolates between the two.

\subsection{The effective action}

We now present an application that motivated this work, namely finding the 
effective action on a stack of M2-branes using AdS/CFT for the particular bulk configurations that we have considered. This 
effective action gives the response of the boundary theory to the action of an external source. Our abelian configuration corresponds
either to a Higgs or a Coulomb phase, depending on the operator that gets a vev. As we have discussed, there are two possible 
meanings to this effective action; the usual theory and the dual one where the on-shell bulk action is identified 
with the boundary effective action. Since the main focus of this paper has been the dual CFT, we will consider 
the latter.

In the case of instantons there is no back-reaction, hence we are ensured that the asymptotic expansion of the 
on-shell action gives the full quantum effective action at large $N$. For a particular solution with the sources set 
to zero this was shown in section 2.6 to be a pure number, $8\pi^2|\l|/g^2$. In the case of non-zero back-reaction, 
since the back-reaction sets in only at order $r^4$, gravity and the matter fields decouple near the boundary and we 
can solve the asymptotic equations in a fixed background. At order $r^4$ the back-reaction sets in as in \eq{grx}, 
but this does not affect the two-point functions. The effective action up to this order 
reduces in the IR to:
\be
S_{\sm{eff}}=\half\int \left(F_{ij}[A]\Box^{-1/2}F_{ij}[A]+\th \e^{ijk}A_i\pa_jA_k\right)~.
\ee
$A$ is the one-point function of the gauge field as a response to the source $J$. The above is in fact the 
on-shell action for the Dirichlet boundary problem, and it is the generic behavior that we found 
under massive deformations: the theory flows towards the Dirichlet problem in the IR. We get a similar result if we start with the massive case \eq{mons}:
\be
S_{\sm{eff}}=\half\int \left(F_{ij}[A]\Box^{-1/2}F_{ij}[A]+\th m\e^{ijk}A_i\pa_j\Box^{-1/2}A_k\right)~.
\ee
We remark here that for zero theta angle, the above is the effective action of three-dimensional QED at large $N_f$. As we 
noted in the previous section, there is another point where our results seem to agree with QED. The naive UV dimension of the 
current is 3/2 rather than 2; the two-point function of such an operator in momentum space would be a contact term; this is 
exactly what we find in \eq{2ptUV}. As discussed, if such a term is to be interpreted as an operator then it must be a non-primary operator. The present computation also predicts that the UV and IR couplings are 
each other's S-duals\footnote{The existence of the two-point function with the operator of dimension 1/2 suggests that in 
the UV the gauge field is dualized to a scalar: $F_i=\pa_i\f$, in agreement with \cite{BKW}.}. The subleading correction is 
the two-point function of the dual gauge field.

\subsection{Non-abelian case}

Due to interactions, classical electric-magnetic duality does not straightforwardly generalize to non-abelian gauge theories
\cite{DT} (but see \cite{DS} where electric-magnetic duality is achieved at the cubic level). The key point is that, as 
explained earlier, a formal replacement $E\leftrightarrow\pm B$ in the Lagrangian does not suffice. Duality has to be
realized at the level of the gauge potential. Duality in non-abelian theories involves quantum effects in an essential way,
which goes beyond the scope of this work. We will suffice with a few comments and suggestions for further work.

Instanton configurations that minimize the energy will satisfy:
\be\label{SDnA}
E_i^a=F_{ri}^a=-\half\,\ve\,\e_{ijk}F^{a\,jk}~.
\ee
Restricted to the boundary, this gives again the same boundary condition as before. 
If in addition we also choose the boundary condition $E_i^a=mA_i^a$, we get the non-abelian version of the self-dual
equation:
\be
A_i^a={1\over2m}\,\e_{ijk}F_{jk}^a~.
\ee
This is the non-abelian generalization of \cite{DJ}.
We have checked that usual 't Hooft instantons do not satisfy this condition but a modification of it. It would be interesting
to see what boundary action gives rise to those solutions.

In section \ref{phys} we outlined a method to compute the two-point function at any of the points in the diagram in Figure \ref{diagram}. In particular, to compute the $S$-dual two-point functions we use electric rather than magnetic bulk sources, that is we couple the source to the gauge field rather than to the electric field. This method can be generalized to the non-abelian case. That is why we expect that some of our results may easily generalize to that case.

As we have noted, the particular compactification we have used has a non-abelian extension where the gauge symmetry is $SU(2)$,
with ${\cal N}=2$ supersymmetry. As mentioned in the introduction, it is an interesting question whether this theory still
has some sort of electric-magnetic duality and what is the holographic image of it. One may be able to shed light on this
question by analyzing instanton solutions considered above. These have $T_{\m\n}=0$, and therefore gravity and the gauge
fields basically decouple. It would be interesting to see whether the theory around such configurations exhibits definite
electric-magnetic properties of the type in \cite{SW}.

\section{Discussion}

The off-shell bulk electromagnetic action is invariant under electric-magnetic duality up to boundary terms. In this paper we 
have analyzed the way these terms transform for arbitrary choice of boundary conditions. The fact that the action is 
invariant off-shell implies that the boundary conditions transform as well. In the simplest example, Neumann and 
Dirichlet boundary conditions for the gauge field are interchanged by $S$-duality. Classically, there is a one-parameter
family of deformations connected by bulk duality. Quantum mechanically only a discrete subgroup survives. It would be interesting
to check whether the latter statement also follows from quantization of Chern-Simons couplings. Operators of higher dimension
are also mapped into each other by electric-magnetic transformations. Again, quantum mechanically the coefficients within a
particular $SL(2,{\mathbb{Z}})$ orbit do not appear in arbitrary but only in specific  combinations. In this way, 
electric-magnetic duality gives a way of probing the moduli space of deformations of effective CFT actions for operators of given dimension. This agrees with interesting experimental results in quantum Hall systems \cite{BD1,BD2}. In the massive case, $S$-duality can change dimension.

We presented a bulk computation of the finite renormalization of dimensions of operators under RG flow of the 
deformed three-dimensional SCFT. The conserved current has IR dimension 2 but the expansion of its two-point function
contains an $S$-dual dimension 1 gauge field in the UV. In other words, the 2-point function contains a
contribution of the 2-point function of the dual current. This is in agreement
with the field theory expectations \cite{tassos}. We also discussed the similarity of the effective action and the flow with those of 2+1 dimensional QED at large $N_f$. It may turn out that the high-energy limit of this theory is relevant to the SCFT in a phase where the non-abelian gauge symmetry is broken, or the theory has become classical due to the large $N$ limit. We notice, however, that whenever we have bulk instanton solutions we will get Chern-Simons terms on the boundary, which as stressed in \cite{schwarz} are not present in SYM theory.
It would be interesting to study this further.

The form of the $S$-transformation worked out in \cite{Wittensl2z} depends on the choice of boundary 
terms. As it turns out, it corresponds to pure Dirichlet and Neumann boundary conditions. In this paper we wrote down the
$S$-generator for arbitrary boundary conditions.
We found that generic choices of boundary conditions are actually invariant under electric-magnetic duality. A particularly 
interesting case is the self-dual massive bulk solution. In this case, the boundary conditions do not entirely determine the
solution, and there is a single remaining boundary degree of freedom. This degree of freedom in fact corresponds to the 
self-dual topologically massive theory in three dimensions \cite{PTvN,DJ}. This should shed some new light on the meaning of 
self-duality of topologically massive theories in three dimensions: it in fact corresponds to electric-magnetic duality in the bulk of 
AdS$_4$! It would be interesting to extend this analysis to the non-abelian case, and we offered some thoughts to analyze
this question. In particular, it seems that the method of doing AdS/CFT with both electric and magnetic sources should work also in that case. One could then test to what extent the effective non-abelian CFT's have dual properties.\\
\\
{\it Note added}: While this paper was being completed, \cite{herzog} appeared, where some related questions are studied.

\section*{Acknowledgements}

\addcontentsline{toc}{section}{Acknowledgements}

We thank Stanley Deser, Marc Henneaux, Don Marolf, and Ioannis Papadimitriou, for useful comments and discussions, and especially Tassos Petkou for collaboration during the early stages of this work.

\appendix

\section{Another proof of $S$-duality}\label{2nds}

We now give the second proof, which relies on integrating out some fields (see also \cite{wittenabelian}). The equations of motion of \eq{1orderwcoupling} are:
\bea\label{eomtiE}
\d_E&:&\pa_rA_i-\pa_iA_r+g^2\,{\sf E}_i+{g^2\th\over4\pi^2}\,F_i=0~,\nn
\d_A&:&-\pa_rE+{1\over g^2}\left(1+{g^4\th^2\over(4\pi^2)^2}\right)*\dd F+{g^2\th\over4\pi^2}\,*\dd\sf E=0~,\nn
\d_{A_r}&:&\pa^i{\sf E}_i=0~.
\eea

We first integrate out $\sf E$ using its equation of motion and the fact that it is conserved, getting the usual second-order action for
the boundary component of the gauge field $A$:
\be\label{Aaction}
S[A;g,\th]=\int\dd^4x\left(-{1\over 2g^2}\,(\pa_rA)^2+{1\over2g^2}\,F^2-{\th\over4\pi^2}\,(\pa_rA)F\right)~.
\ee
This is the usual second order action \eq{actionwtheta}.

Next we use the equation of the gauge field to integrate out $A$ instead. We make use of conservation of $\sf E$ to introduce a gauge field
$\sf v$:
\bea
\sf E&=&*\dd \sf v\nn
\sf v&=&v-{\th\over4\pi^2}A~.
\eea

Then the second equation in \eq{eomtiE} is integrated to:
\be
F={g^2\over1+{g^4\th^2\over(4\pi^2)^2}}\,\pa_r{\sf v}-{g^4{\th\over4\pi^2}\over1+{g^4\th^2\over(4\pi^2)^2}}\,{\sf E}~.
\ee
Observe that this is nothing but the $S$-dual of the equation of motion for $\sf E$! We get:
\be\label{Sd}
S[{\sf E}]=-\int\dd^4x\left(-\half{g^2\over1+{g^4\th^2\over(4\pi^2)^2}}\,(\pa_r{\sf v})^2+ \half{g^2\over1+{g^4\th^2\over(4\pi^2)^2}}\,{\sf E}^2+
{g^4{\th\over4\pi^2}\over1+{g^4\th^2\over(4\pi^2)^2}}\,(\pa_r{\sf v}){\sf E}\right)~.
\ee
This is the $S$-dual of \eq{Aaction}! 

In conclusion, the gauge fields $A$ and ${\sf v}$ are each other's $S$-duals: integrating out $\sf v$ we get the usual second form of the
action; integrating out $A$ gives its $S$-dual version where the gauge field is $\sf v$. The first order form of the action, on the other hand,
\eq{1orderwcoupling} or equivalently \eq{1storderform}, interpolate between both. In particular, \eq{1orderwcoupling} has manifest $S$-duality.

The on-shell action now is:
\be
S_{\sm{on-shell}}=-\half\int\dd^3x\,\sf E A~,
\ee
which is obviously the same as \eq{onshellwth}. Notice however that in terms of the redefined $\sf E$, the on-shell action takes the same
form as the on-shell action without the $\th$-term.

\section{Regularity of the solutions}\label{regularityapp}

In this appendix we study the regularity condition on bulk solutions and give some other formulas that were used in the main text.

\subsection{Gauge invariance}

We first discuss gauge invariance.
The longitudinal, pure gauge part $\vf=-{i\over p^2}\,p^iA_i$ is not fixed by the equations 
of motion. 
\bea\label{decomp}
A_i(r,x)&=&A_i^{\tiny{\mbox{T}}}+\pa_i\vf\nn
A_r(r,x)&=&\pa_r\vf\nn
F_{ri}(r,x)&=&\pa_rA_i^{\tiny{\mbox{T}}}~.
\eea
From this, the radial gauge $A_r=0$ is simply $\pa_r\vf=0$, leaving the residual 
$r$-independent gauge freedom $\vf(x)$. $\vf(x)$ realizes the gauge 
symmetry of the dual CFT, where the background field has been gauged and a dual current is constructed from the
dual gauge field $A$. By construction, $A_i^{\tiny{\mbox{T}}}$ and $f_i$ 
are invariant under such transformations.
 
\subsection{Regularity in two different bulk gauges}
As the topological term does not effect equations of motion, we will take the Maxwell action without $\theta$-term
\be\label{emaction}
S_{\sm{bulk}}(A)={1\over2}\int d^4x\sqrt{g}\,F_{\m\n}F^{\m\n}
\ee

We will study two different gauges. The first is $\nabla_{\m}A^{\m}=0$, where the bulk equations are 
\be\label{emeq}
\nabla_{\m}F^{\m\n}=0,\quad \nabla_{\m}A^{\m}=0
\ee
Making use of the following identities
\bea
&& \partial_r^2-{2\over r}\partial_r+{2\over r^2}=(\partial_r-{1\over r})^2\nonumber\\
 &&(\partial_r-{1\over r})^n(r\phi)=r\partial_r^n\phi
 \eea
and denoting ${\cal A}={A_0(r,\bar x)/r}$, we rewrite the equations 
\bea\label{emeq2}
 (\partial_r^2+\bar\partial^2){\cal A}&=&0\nonumber\\
 (\partial_r^2+\bar\partial^2)A_i&=&2\partial_i{\cal A}
 \eea
which are solved by
\bea\label{sol}
 {\cal A}(r,\bar p)&=& {\cal A}_0(\bar p)\cosh(|\bar p|r)+{1\over|\bar p|}{\cal A}_1(\bar p)\sinh(|\bar p|r)\nn
  A_i(r,\bar p)&=&A_i^{(0)}\cosh(|\bar p|r)+{1\over|\bar p|}{A}_i^{(1)}\sinh(|\bar p|r)\nn
               &+& ir{\bar p_i\over|\bar p|}\left({\cal A}_0\sinh(|\bar p|r)+{1\over|\bar p|}{\cal A}_1\cosh(|\bar p|r)\right)\nn
\eea
The gauge fixing $\nabla_{\m}A^{\m}=0$ implies
\be\label{aux}
 {\cal A}_0(\bar p)=i\bar p_iA_i^{(0)}(\bar p),\quad  {\cal A}_1(\bar p)=i\bar p_iA_i^{(1)}(\bar p)
\ee
The two boundary functions $A_i^{(0)},A_i^{(1)}$ completely determine the bulk solutions. 

Near the boundary $r\!\rightarrow\!0$ we find
 \bea\label{r0}
 &&A_i(r,\bar p)=A_i^{(0)}(\bar p)+r\,\Pi_{ij}A_j^{(1)}(\bar p)+O(r^2)\nonumber\\
 &&{E}_i=F_{0i}=\Pi_{ij}A_j^{(1)}(\bar p)+r\,\bar p_i\bar p_jA_j^{(0)}+O(r^2)
 \eea
 where
\be\label{proj}
\Pi_{ij}=\delta_{ij}-{p_ip_j\over p^2}
\ee
By construction, this matrix has a null eigenvector $p_i$. It's clear that $f_i=E_i^{(0)}$ is the transverse part of $A_i^{(1)}$.
 In position space
 \be\label{Air0}
 A_i(r,\bar x)=A_i^{(0)}(\bar x)+r{E}_i^{(0)}(\bar x)+...
 \ee
 where
 \be
f_i(\bar x)=E_i^{(0)}(\bar x)=\int d^3\bar y {A_j^{(0)}(\bar y)\over |\bar x-\bar y|^4}\left(\delta_{ij}-{(\bar x_i-\bar y_i)(\bar x_j-\bar y_j)\over |\bar x-\bar y|^2}\right)
\ee
 
 To find regular solutions, we expand the general solution (\ref{sol})
 \bea\label{regular}
 {A}_i&=&{1\over2}\left({A}_i^{(0)}+{1\over|\bar p|}{A}_i^{(1)}\right)e^{|\bar p|r}+{1\over2}\left({A}_i^{(0)}-{1\over|\bar p|}{A}_i^{(1)}\right)e^{-|\bar p|r}\nonumber\\
 &&+i{\bar p_i\over|\bar p|}\left[{1\over2}\left({\cal A}_0+{1\over|\bar p|}{\cal A}_1\right)re^{|\bar p|r}-{1\over2}\left({\cal A}_0-{1\over|\bar p|}{\cal A}_1\right)re^{-|\bar p|r}\right]
 \eea
To remove the divergent terms, it's enough to require 
 \be\label{sol1}
 {A}_i^{(0)}(\bar p)+{1\over|\bar p|}{A}_i^{(1)}(\bar p)=0
 \ee

Next we study the radial gauge $A_r=0$, with bulk equations
\be
\nabla_{\m}F^{\m\n}=0,\quad A_{r}=0
\ee
These reduce to 
\be\label{rgauge}
(\partial_r^2+\bar\partial^2)A_i(r,\bar x)=\partial_i\partial_jA_j(r,\bar x),\quad \partial_r\partial_iA_i=0
\ee
which is simply a rewriting of (\ref{emeq2}) if one replaces ${\cal A}$ by $\partial_jA_j$. Notice this replacement is consistent with (\ref{aux}). The second equation in (\ref{rgauge}) imposes the Gauss law $\partial^iE_i=0$. 

In summary, we found regularity requires in either gauge 
\be\label{Af}
\Pi_{ij}A_j(p)+{1\over{|p|}}f_i(p)=0
\ee
in terms of the boundary fields $A_i$ and $f_i$. In momentum space, we denote the transverse part of the boundary field
\be\label{transverse}
A_i^{\tiny{\mbox{T}}}(p)=\Pi_{ij}A_j(p)\,.
\ee
The regularity condition is then
\be\label{tAf}
A_i^{\tiny{\mbox{T}}}(p)+{1\over{|p|}}f_i(p)=0
\ee

\subsection{Regularity without gauge fixing}
We take advantage of the conformal invariance of the Maxwell equations. In Poincare coordinates they take the form
\be
\nabla_{\m}F^{\m\n}=0~,
\ee
where the metric is the flat Euclidean metric. These reduce to
\bea\label{rgauge}
\left(\partial_r^2+\Box\right)A_i(r,\bar x)&=&\pa_i\pa^jA_j+\pa_i\pa_rA_r\nn
\Box A_r&=&\pa_r\pa^jA_j~,
\eea
where $\Box=\pa^j\pa_j$. The latter is the conservation equation for the electric field $E_i=F_{ri}$, $\pa^iE_i=0$.

To separate the different degrees of freedom, we decompose the gauge field in the following way
\bea\label{varsep}
A_i&=&\pa_i\vf+A_i^{\tiny{\mbox{T}}}\nn
\Box\vf&=&\pa^iA_i
\eea
Clearly, $A^{\tiny{\mbox{T}}}$ is the transverse part of the gauge field $\pa^iA_i^{\tiny{\mbox{T}}}=0$,
which carries the physical polarizations. Indeed, $A_i^{\tiny{\mbox{T}}}=A_i-\Box^{-1}\pa_i\pa^jA_j$
is invariant under gauge transformations. We find
\be
E_i=\pa_rA_i^{\tiny{\mbox{T}}}~.
\ee
In momentum space, we have
\be
A_i^{\tiny{\mbox{T}}}(r,p)=\Pi_{ij}A_j(r,p)\,.
\ee

The longitudinal, pure gauge part of the gauge field is described by $\vf$. In momentum space, $\vf=-{i\over p^2}\,p^jA_j$.

Fill the above decomposition into the equations of motion. We get
\be\label{Ar}
A_r=\pa_r \vf~,
\ee
and so $A_r$ is not determined, since $\vf$ isn't. The remaining equation of motion reads
\be
\left(\pa_r^2+\Box\right)A_i^{\tiny{\mbox{T}}}=0~.
\ee
This is solved by
\be
A_i^{\tiny{\mbox{T}}}(r,\bar p)=A_{(0)i}^{\tiny{\mbox{T}}}\cosh(|\bar p|r)+{1\over|\bar p|}\,A_{i(1)}^{\tiny{\mbox{T}}}\sinh(|\bar p|r)~.
\ee
We recognize this as the homogeneous piece of the solution in (\ref{sol}), as the other piece there is pure gauge.

We will now impose regularity of the solution. Rewrite the solution as
\be\label{solu}
A_i^{\tiny{\mbox{T}}}={1\over2}\left(A_{i(0)}^{\tiny{\mbox{T}}}+{1\over|\bar p|}\,A_{i(1)}^{\tiny{\mbox{T}}}\right)e^{|\bar p|r}+{1\over2}\left(A_{i(0)}^{\tiny{\mbox{T}}} -{1\over|\bar p|}\,A_{i(1)}^{\tiny{\mbox{T}}}\right)e^{-|\bar p|r}~.
\ee
As before, regularity at $r\rightarrow\infty$ requires
\bea\label{sol11}
A_{i(0)}^{\tiny{\mbox{T}}}(\bar p)+{1\over|\bar p|}\,A_{i(1)}^{\tiny{\mbox{T}}}(\bar p)&=&0~.
\eea
We can rewrite this in terms of the full gauge field as:
\bea\label{regularity1}
\Pi_{ij}\left(A_{(0)j}+{1\over|p|}\,A_{(1)j}\right)&=&0\nn
A_{(0)i}+{1\over|p|}\,A_{(1)i}&=&ip_i\left(\vf_{(0)}+{1\over|p|}\,\vf_{(1)}\right)~,
\eea
where we expanded $\vf$ in the usual way. Hence, regularity relates $A_{(0)}$ and $A_{(1)}$ up to an arbitrary gauge transformation.

The self-duality boundary condition \eq{sdbc} can be rewritten as
\be
A_{(1)i}^{\tiny{\mbox{T}}}=i\ve\,\e_{ijk}p_jA_{(0)k}
\ee
which is again rewritten as
\be
\Pi_{ij}\left(A_{(1)j}-i\ve\,\e_{jkl}p_kA_{(0)l}\right)=0~.
\ee
Combining regularity \eq{regularity1} and self-duality, we get
\be
\Pi_{ij}\left(A_{(0)j}+{i\ve\over|p|}\,\e_{jkl}p_kA_{(0)l}\right)=0~.
\ee
Again, the combination between brackets is set to zero up to a gauge transformation.

Imposing the gauge $A_r=0$ sets $\pa_r\pa^jA_j=0$, which at lowest order simply sets the longitudinal part of $A_{(1)}$ to zero, $\vf_{(1)}=0$. The longitudinal part of $A_{(0)}$ is left undetermined. This corresponds to the residual gauge symmetry of \eq{rgauge} corresponding to $r$-independent gauge transformations. These act purely on the boundary value of the gauge field. In that case, the gauge invariant form of the regularity condition is
\be
A_{(0)i}(p)+{1\over|p|}\,A_{(1)i}=ip_i\vf_{(0)}~.
\ee
Combined with self-duality thus gives
\be
A_{(0)i}+{i\ve\over|p|}\,\e_{ijk}p_jA_{(0)k}=ip_i\vf_{(0)}~.
\ee

\subsection{Some used identities}\label{ids}

\bea\label{id4}
\Pi^2&=&\Pi\nn
*\dd F&=&*\dd*\dd A=-t\Box\Pi A\nn
\e_{ikm}\e_{jlm}\pa_k\pa_l&=&t\,\Box\Pi_{ij}\nn
\Pi*\dd&=&*\dd\Pi=*\dd\nn
(*\dd*\dd)_{ij}&=&-M^2\Pi_{ij}=-t\Box\Pi A\nn
*\dd(M\Pi)^{-1}*\dd(M\Pi)^{-1}*\dd&=&-*\dd\nn
\left(\Pi-{\th\over M}\,*\dd\right)^{-1}&=&{\Pi+{\th\over M}\,*\dd\over1+\th^2}~.
\eea

Using regularity, we get
\be
\Pi_{ij}v_j={it\over M^2}\,\e_{ijk}p_jf_k
\ee
so
\be
v_if_i=t\,A_iF_i~.
\ee

\section{Solution of the boundary self-dual equation}\label{solutionSD}

\subsection{General regular solution}

Let us consider the general action \eq{generalaction}. We rewrite its equations of motion \eq{generaleom}, combined with the regularity condition \eq{gaugeinvreg}:
\be\label{reg}
f=-M\,\Pi A~,
\ee
and obtain equations for regular solutions
\bea\label{regeom}
(\a-\m M)\,F+(1-\g)M\,\Pi\cdot A+\d\,*\dd F&=&\ti J\nn
-\b M\,A+(\g-\e M)\,F+\m\,*\dd F&=&J~.
\eea

We now compute the on-shell action using regularity:
\bea\label{osefa}
S_{\sm{on-shell}}&=&\int\dd^3x\left(\left(\half\a-\m M\right)AF-\half\b Mv\Pi A+(\half\e M^2-M\g+\half M)\,A\Pi A+\half\,\d F^2\right.\nn
&-&\left.A(\ti J-M\Pi{\cal J})\right)~.
\eea
We can partial integrate to find
\bea
-{1\over2}\b\int d^3x\, v_iM\Pi_{ij}A_j={t\over2}\b\int d^3x\,A_iF_i~.
\eea
and also $\int F^2=t\int M^2A\Pi A$. The effective action is finally
\bea\label{osefa1}
S_{\sm{on-shell}}&=&\int\dd^3x\left(\left(\half\a-\m M+\half\b\right)AF+\left(\half\e M^2-M\g+\half M+\half\,t\d\,M^2\right)\,A\Pi A\right.\nn
&-&\left.A(\ti J-M\Pi{\cal J})\right)~.
\eea
This expression, although not yet entirely explicit because it still depends on $A$, and was used in the main text.

We now solve the equations of motion in general, for regular Euclidean bulk solutions. Using regularity in the equations of motion \eq{generaleom}, after some algebra we get:
\bea
-M(\b+\m tM)\Pi A+(\g-\e M)\,F&=&*\dd J\nn
-M(\g-1+\d tM)\Pi A+(\a-\m M)\,F&=&*\dd\ti J~.
\eea
This has now become purely a system of equations for $A$. Since there is but one independent equation for $A$, both equations
have to be consistent. Without loss of generality, let us take this to be the first equation above, which we rewrite as
\be\label{genmassive}
-a \Pi A+bF=J~,
\ee
After some algebra, we can rewrite this as a massive equation for $A$, as we did before:
\be\label{generalmassive}
\left(a^2-t\,b^2M^2\right)\Pi A=j~,
\ee
where we defined a new conserved current
\be
j=-b*\dd J-a\,J~.
\ee
The on-shell effective action can now be written entirely in terms of $j$.

\subsection{General topologically massive solution}

In this appendix we explicitly solve the equations of motion of the massive topological boundary theory.

We showed in Appendix \ref{regularityapp} that the self-duality solution combined with regularity give the following equation in momentum space:
\be
A_i^{\mbox{\tiny{T}}}(p)=-{i\ve\over|p|}\,\e_{ijk}p_jA_k^{\mbox{\tiny{T}}}(p)~.
\ee
The solution is expanded in a basis of polarization vectors $p_i$, $p_i^*$, $k_i$, where
$k\cdot p=k\cdot p^*=0$. Choosing coordinates $p_i=(p_0,p_1,p_2)$, we have $p_i^*=(-p_0,p_1,p_2)$, and we can choose for example $k_i=(0,-p_1,p_2)$. Expanding
\be\label{Asolution}
A_i^{\mbox{\tiny{T}}}(p)={p_i^*\over|p|}\,f(p)+{k_i\over|p|}\,g(p)~,
\ee
we find the constraints
\bea\label{B3}
p\cdot p^*&=&0\nn
g(p)&=&-{4i\ve p_1p_2\over\sqrt{2}p_0^2}\,{\mbox{sgn}}(p_0)\,f(p)~.
\eea
Imposing the massive boundary condition \eq{dtbc} simply adds to this the constraint $p^2=m^2$ to the above solution.

It will be useful to have the general solution of the self-dual theory
\be\label{AeA}
A_i={1\over m}\,\e_{ijk}\pa_jA_k~.
\ee
in position space. First of all, \eq{AeA} automatically implies the gauge fixing condition 
\be\label{gf}
\pa^iA_i=0~.
\ee
We can rewrite \eq{AeA} as follows:
\be
\left(\Box+m^2\right)A_i=0~,
\ee
where $\Box$ is with respect to the flat Euclidean 3-dimensional metric. The general solution of this is best written in terms of Bessel functions of the first kind:
\be
A_i(R,\Omega_2)={1\over\sqrt{R}}\sum_{lm}\left(a_{i(1)lm}\,J_{l+\half}(mr)+a_{i(2)lm}\,J_{-l-\half}(mr)\right) Y_{lm}(\Omega_2)
\ee
in polar coordinates $(R,\Omega_2)$, for some constants $a_{i(1)}$ and $a_{i(2)}$. These constants are however not all arbitrary. Indeed, it is easy to show that the theory has a single massive degree of freedom, described by a {\it single} pair of polarizations $a_{i(1)}$, $a_{i(2)}$. One of the three $A_i$'s is solved for from \eq{AeA}. The remaining one is solved from \eq{gf}.

Solving the equation in momentum space as in \eq{Asolution}, we find
\bea
p\cdot p^*&=&0\nn
g(p)&=&{4\sqrt{2}i\over m}\,{\mbox{sgn}}(p_0)\,p_1p_2\,f(p)~.
\eea

\subsection{Boundary terms for the self-dual problem}\label{sdactions}

As remarked in the main text, because of the invariant of the self-dual solutions, self-dual boundary conditions do not lead to unique boundary terms. We give here two other actions that lead to the same boundary problem:
\bea\label{moresdac}
S_{\sm{bdy}}[A,f]&=&\int\dd^3x\left(Af+{\ve\over2m}\,F^2-{1\over m}\,fF\right)\nn
S_{\sm{bdy}}[A,f]&=&\int\dd^3x\left(Af+\half\,\ve\,AF-{1\over m}\,fF\right)~.
\eea
It would be interesting to check that these are connected by an $SO(1,1)$ transformation.

We finally also give the values of the constants in \eq{generaleom} and \eq{generalmassiveeq}:
\bea\label{a123}
a_1&=&\b\d-(\g-1)\m\nn
a_2&=&\g\d-\a\m\nn
a_3&=&\d\e-\m^2\nn
b_1&=&\b\m-(\g-1)\e\nn
b_2&=&\g\m-\a\e~,
\eea
and
\bea\label{ab}
a&=&{a_3^2\over a_1b_2-a_2b_1}\nn
b&=&{a_3(a_1-b_2)\over a_1b_2-a_2b_1}~,
\eea
both in Euclidean signature.

\subsection{Near-extremal solutions}\label{nearextremal}

A particularly interesting boundary condition is the generalization of \eq{sdbc} to non-extremal solutions where the electric and magnetic field are still aligned:
\be\label{flF}
f=\l\,F~.
\ee
This boundary condition preserves conformal invariance. Combining it with the massive boundary condition \eq{dtbc}, we get the following action:
\be
S_{\sm{bulk}}^{\sm{on-shell}}=\int\dd^3x\left(\g\,Af+\half\,\ve\,AF+{\e\over2}\,f^2\right)~,
\ee
The parameter $\l$ and the mass are given by:
\bea
\l&=&{1\over1-\g}\nn
m&=&-{\g\over\e}~.
\eea
Near extremality, $\l\simeq\ve$, we get that $\g$ is small. $m$ is kept finite by making $\e$ small as well.

\section{Holographic renormalization of abelian gauge fields}\label{holren}

The total Euclidean action is:

\be\label{Euclaction}
S=\int\dd^4x\sqrt{g}\,\left(-{R-2\L\over16\pi G_N}+{1\over4}F_{\m\n}F^{\m\n}\right)-{1\over8\pi G_N}\int\dd^3x\sqrt{\g}\,K+S_{\sm{bdy}}[A]
\ee
where $\L=-3/\ell^2$, $K_{ij}=\half\pa_r\g_{ij}$, and $\g={\ell^2\over r^2}\,g$ where $g$ is the metric \eq{FG}. The gravity equations of motion are
\bea
G_{\m\n}+\L\,g_{\m\n}&=&8\pi G_N\,T_{\m\n}\nn
T_{\m\n}&=&F_{\m\l}F_\n{}^\l-{1\over4}\,g_{\m\n}\,F_{\l\s}F^{\l\s}~.
\eea
For conformally invariant matter, the stress-energy tensor is traceless. Therefore, it is most convenient to write Einstein's equations as follows:
\be\label{Einsteineq}
R_{\m\n}+{d\over\ell^2}\,g_{\m\n}=8\pi G_N\,T_{\m\n}~.
\ee

The Ricci tensor is:
\bea\label{Ricci}
R_{rr}&=&-\half\,\Tr(g^{-1}g'')+{1\over4}\,\Tr(g^{-1}g')^2+{1\over2r}\,\Tr(g^{-1}g')-{d\over r^2}\nn
R_{ri}&=&\half[\nabla^jg_{ij}'-\nabla_i\Tr(g^{-1}g')]\nn
R_{ij}&=&R_{ij}[g]-\half\,g_{ij}''+\half\,(g'g^{-1}g')_{ij}-{1\over4}\,g_{ij}'\,\Tr(g^{-1}g')+{1\over2r}\left(g_{ij}\Tr(g^{-1}g')+(d-1)g'_{ij}\right)-{d\over r^2}\,g_{ij}~,\nn
\eea
where $d$ is the dimension of the boundary. From the fact that the curvature scalar is always constant we get
\be
\Tr(g^{-1}g'')-{d\over r}\,\Tr(g^{-1}g')-{3\over4}\,\Tr(g^{-1}g')^2+{1\over4}(\Tr(g^{-1}g'))^2-R[g]=0~.
\ee

As usual in holographic renormalization, we solve Einstein's equations perturbatively in the distance to the boundary:
\be\label{FGexp}
g(r,x)=g_{(0)}(x)+r^2g_{(2)}(x)+r^3g_{(3)}+r^4g_{(4)}+\cdots
\ee
We find \cite{dHSS}
\be
g_{(2)}=-{1\over d-2}\left(\mbox{Ric}[g_{(0)}]-{1\over2(d-1)}\,g_{(0)}\,R[g_{(0)}]\right)=0~,
\ee
where we used the fact that in our applications we have the boundary condition $g_{(0)ij}=\d_{ij}$. At the next order, from the first and third of \eq{Ricci}-\eq{Einsteineq} we find the equation 
\bea
\Tr g_{(3)}&=&0\nn
(d-3)\,g_{(3)}&=&0~,
\eea
which leaves the traceless part of $g_{(3)}$ undetermined, as it should, since at this order we find the holographic stress-energy tensor \cite{dHSS}:
\be
g_{(3)}={16\pi G_N\over3\ell^2}\,\bra T_{ij}\ket~.
\ee
The second equation then gives
\be
\nabla^jg_{(3)ij}={16\pi G_N\over3\ell^2}\,\t_{ri}|_{r=0}~,
\ee
therefore
\be\label{Wardid}
\nabla^j\bra T_{ij}\ket=f^jF_{ij}~.
\ee
Since $f^i=-\bra O^i_{\Delta=2}\ket$, where $O^i_{\Delta=2}$ is an operator of dimension 2, this is in exact 
agreement with the Ward identity derived in field theory (see formula (4.19) of \cite{BFS}). Indeed, one considers 
the effective action
\be
\d S_{\sm{eff}}=\int\dd^3x\sqrt{g_{(0)}}\left(\half\d g^{ij}_{(0)}\bra T_{ij}\ket+\bra O^i_{\Delta=2}\ket\d 
A_i\right)~.
\ee
Invariance of the action under diffeomorphism then implies \eq{Wardid}. As remarked earlier, in this paper, instead of regarding $O^i_{\Delta=2}$ as an operator, we will regard it as a conserved current.

At the next order, we find
\bea
g_{(4)}&=&-{4\pi G_N\over\ell^2}\left(\t_{ij}-\half\,\d_{ij}\,\Tr\,\t\right)|_{r=0}\nn
&=&-{4\pi G_N\over\ell^2}\left(f_if_j-F_iF_j-{1\over4}\,g_{ij}(f^2-F^2)\right)~.
\eea

Regularity of the matter solution amounted to a relation between $f$ and $A$. When we include gravity, we have to 
demand regularity of the coupled gravity-Maxwell system. This will in particular involve a special choice of the 
boundary stress-energy tensor, which is left undetermined by \eq{Wardid}. We will not pursue this further here, but 
simply assume that such a choice indeed exists and is enough to make the solutions well-behaved at $r=\infty$. 

We now consider the effect of back-reaction on the asymptotic expansion of the on-shell action, which determines the divergent part of the action. The matter part of the action has  been analyzed in the main text. Including 
back-reaction does not change this, as the on-shell action was shown to be a total derivative. For the gravitational 
part of the action we can now use the formula in \cite{dHSS}
\be\label{regaction}
S=\int_{\r\geq\e}\dd\r\,{3\over\r^{5/2}}\,\sqrt{\det g(\r,x)}+{1\over\r^{3/2}}\left( -6\sqrt{\det g(\r,x)}+4\r\pa_\r\sqrt{\det g(\r,x)}\right)\,
\ee
where $\r=r^2$. This is the regulated part of the action. Now it is easy to check that the contribution from $\r=\e$ is
\be
S=-4\e^{-3/2}~.
\ee
This is the leading term near the conformal vacuum. As usual, this is cancelled by adding a covariant counterterm 
$2(1-d)/\e^{d/2}$. 

From \eq{regaction} we also see that $g_{(4)}$ and any higher order pieces only contribute terms that do not contribute to 
two-point functions. On the other hand, there is a potential contribution from $g_{(3)}$, which contains the 
stress-energy tensor. This gives the usual coupling to the metric \eq{Wardid}
\be
\half\int\dd^3x\sqrt{g_{(0)}}\,g^{ij}_{(0)}T_{ij}~,
\ee
but since with our choice of $g_{(0)}$ the boundary stress-energy tensor is traceless, this vanishes. Therefore, 
only the contribution from the matter part of the action is present in the two-point functions. In fact, 
there will be additional contributions since we will be considering a problem with generalized boundary conditions.

\section{Analysis of supersymmetric boundary conditions}\label{susyapp}

The bosonic fields we discuss here are the bosonic fields of the four dimensional gauged $\CN=2$ supergravity \cite{Freedman:1976aw}. The gauge symmetry is $SO(2)=U(1)$ and can be embedded in the $SO(5)^V$ truncation of the gauged maximal $SO(8)$ supergravity \cite{Romans:1983qi} where all scalar fields are projected out.  The full field contents are the graviton, two real gravitini $\psi_{\m}^i,\,i=1,2$ and the abelian gauge field, which form the $\CN=2$ gravity multiplet. The complete action with the two real gravitini combined into $\psi_\m=\psi^1_\m+i\psi^2_\m$ reads (setting $\kappa=1$) \cite{Caldarelli:2003pb}
\bear
S&=&\int\dd^4x\, \sqrt{g}\left(-{R\over4}+{\L\over2}+{1\over4}F_{\m\n}F^{\m\n}+\half \bar\psi_\m\g^{\m\n\rho}D_\n\psi_{\rho}-{1\over2\ell}\bar\psi_\m\g^{\m\n}\psi_\n \right.\nonumber\\
&&\left.+{i\over4}(F^{\m\n}-\half {\rm{Im}}(\bar\psi^\m\psi^\n))\bar\psi_{\rho}\g_{[\m}\g^{\rho\sigma}\g_{\n]}\psi_{\sigma}\right)
\eear
where
\bear
 D_\m&=&\nabla_\m-{i\over\ell}A_\m\nn
\nabla_\m&=&\pa_\m+{1\over4}\lambda_\m{}^{ab}\g_{ab}
\eear
and the spin connection is given by 
\bear
\lambda_{\m ab}&=&\O_{\m ab}-\O_{\m ba}-\O_{ab\m}\nn
\O_{\m\n}{}^a&=&\pa_{[\m}e^a_{\n]}-\half{\rm{Re}}(\bar\psi_\m\g^a\psi_\n)
\eear
Define the super-covariantized field strength 
\be
\hat F_{\m\n}=F_{\m\n}-{\rm Im}(\bar\psi_\m\psi_\n)
\ee
and the super-covariant derivative
\be
{\cal D}_\m=D_\m+{1\over2\ell}\g_\m+{i\over4}\hat F_{ab}\g^{ab}\g_\m
\ee
we can write the supersymmetry transformations as 
\bea\label{susy}
\d e^a_\m&=&{\rm Re}(\bar\epsilon\g^a\psi_\m)\nn
\d A_\m&=&{\rm Im}(\bar\epsilon\psi_{\m})\nn
\d\psi_\m&=&{\cal D}_\m\epsilon
\eea
where $\epsilon=\epsilon^1+i\epsilon^2$. The $U(1)$ gauge transformation is 
\be
\d A_\m=\pa_\m\a,\quad\d\psi_\m=-i\a\psi_\m
\ee

To determine the supersymmetric boundary conditions, we take the approach recently studied in \cite{Hollands:2006zu}. First, we look for boundary conditions on the bosonic and fermionic fields which render the symplectic flux through the boundary vanishing. As we showed in Appendix \ref{holren}, we can expand close to the boundary $r=0$
\be
g(r,x)=\d+r^3g_{(3)}+r^4g_{(4)}+\cdots
\ee
which gives the boundary condition on metric, esp. $\d g=O(r^3)$. The gauge field contributes to the symplectic current
\bea
*\o_r[A]=-\d_1A^\m \d_2(F_{r\m}+{i\over2}\bar\psi_{\rho}\g_{[r}\g^{\rho\sigma}\g_{\m]}\psi_{\sigma})+\d_2A^\m \d_1(F_{r\m}+{i\over2}\bar\psi_{\rho}\g_{[r}\g^{\rho\sigma}\g_{\m]}\psi_{\sigma})
\eea
Notice that the pure gauge field contribution vanishes provided the matrices in \eq{sym} are symmetric. The fermion bilinear terms contribute to the symplectic flux together with those involving only the gravitini, which we now discuss.

 Compared to the minimal $\CN=1$ theory, the $\CN=2$ action contains in addition coupling to the gravi-photon field and extra fermion interactions. However, it's straightforward to show that these couplings do not contribute to the gravitini symplectic current. Hence we find the symplectic current from the spin-2 and 3/2 fields \cite{Hollands:2006zu}
 \bea
*\o^r[g,\psi] &=& r^3P^{ijklm r} 
[\delta_1 g_{ij} \delta_2 \lambda_{k\, lm} - 
(1 \leftrightarrow 2)]
\nonumber\\
&+& r^{-2}\epsilon^{rijk} [\delta_1 \ti{\bar \psi_i}
\g_5 \ti\g_j \d_2 \ti\psi_k + \half \d_1 g_j{}^m 
\ti{\bar \psi_i} \g_5 \ti\g_m \d_2 \ti\psi_k
-(1 \leftrightarrow 2)] \nn
&-&r{i\over2}g^{rr}[\d_1A^m \d_2(\ti{\bar \psi_{\rho}}\ti\g_{[r}\ti\g^{\rho\sigma}\ti\g_{m]}\ti\psi_{\sigma})-(1 \leftrightarrow 2)]
 \eea
 where
 \bea
 P^{\alpha\beta\gamma\tau\sigma\mu} = -\frac{1}{2}
g^{\alpha\beta}g^{\gamma\tau}g^{\mu\sigma}
+ \frac{1}{2}  
g^{\alpha\mu}g^{\gamma\tau}g^{\beta\sigma}
+ \frac{1}{2}  
g^{\alpha\gamma}g^{\beta\tau}g^{\mu\sigma}~.
 \eea
The symplectic from is written in terms of rescaled metric $g_{ij}$ and fermions $\ti\psi_k=r^{1/2}\psi_k$. Now it can be seen that the graviton contribution to the symplectic flux vanishes 
\be
\int_{r=0}\o[g]=0
\ee
where the dual symplectic current
\be
\o[g]=\epsilon_{ijk\,r}*\o^r\,\dd x^i\wedge\dd x^j\wedge\dd x^k=O(r^2)~.
\ee 
In the last equation we used $\d g=r^3\d g_{(3)}$. We also see that the gravitini contribution 
\be\label{symptini}
\int_{r=0}\o[\psi]=\int_{r=0}\left({1\over r^2}{\delta_1 \ti{\bar \psi_{i}}} \ti\g_j \g_5 \d_2 \ti\psi_{k} -{i\over2r}\ti\epsilon_{ijk}\d_1A^m \d_2(\ti{\bar \psi_{\rho}}\ti\g_{[r}\ti\g^{\rho\sigma}\ti\g_{m]}\ti\psi_{\sigma})\right)\, \dd x^i \wedge \dd x^j \wedge \dd x^k
\ee
is finite but non-vanishing if $\ti\psi_\m=O(r)$.  In order that the symplectic flux vanishes, we need $\ti\psi_i=O(r^2)$ and $\ti\psi_r=O(r)$. In terms of $\psi_\m$ we impose \be\label{bcvitini}\psi_r=O(r^{1/2}),\quad \psi_i=O(r^{3/2})~.\ee

Having found the boundary conditions consistent with quantization, it's interesting to see whether supersymmetry is preserved asymptotically. This is equivalent to that supersymmetry transformations keep the symplectic current invariant on the boundary.  Plugging \eq{bcvitini} into \eq{susy} leads to
\be
 \d_{\epsilon} A_r=O(r^{1/2}),\quad \d_{\epsilon} A_i=O(r^{3/2})~.
\ee
From this we see that $\d_{\epsilon} A_i$ contributes a vanishing term to the gauge field symplectic current. Evaluating the transformation of $g_{\m\n}$ is slightly more complicated, but one can show \cite{Hollands:2006zu} that up to an infinitesimal diffeomorphism the susy variation takes the form $\d_{\epsilon} g=O(r^2)$. The infinitesimal diffeomorphism $\pounds_{V_{\epsilon}}$ changes $g_{\m\n}$ at order $O(r)$ and simply signals a departure from the Gaussian normal gauge. In practice, one can take the supersymmetry generator to be $s_{\epsilon}-\!\pounds_V$. It now remains to show that the $\psi_\m$ boundary conditions are also preserved by supersymmetry. 

For convenience of notation we define the operator ${\cal D}_0={\cal D}|_{A=0}$. Using the result of Appendix \ref{holren} the stress tensor $T_{\m\n}=O(r^2)$, one can show following \cite{Hollands:2006zu} that for $\eta$ a ${\rm AdS_4}$ killing spinor 
\be
\d_{\eta}^0\psi_\m={\cal D}_{0\,\m}\eta=O(r^{3/2})~.
\ee
What remains is to find the behavior of 
\be
\d_{\eta}^A\psi_\m=({\cal D}_\m-{\cal D}_{0\,\m})\eta=-\left({i\over\ell}A_\m-{i\over4}\g^{\a\b}\g_\m F_{\a\b}\right)\eta~.
\ee
For generic values of $A$ and $F$, the first term is order $O(r^0)$ while the second is $O(r)$. Moreover since $A_\m$ is a $U(1)$ gauge field, both $A_\m$ and $F_{\a\b}$ are just numbers and the two terms are linearly independent due to the $\g$-matrix.  Recall $\ti\psi_\m=r^{1/2}\psi_\m$ and \eq{symptini} we find $A_\m=O(r)$ in order that the symplectic current is finite at the boundary. This suggests the only supersymmetric boundary condition for the gauge field is the Dirichlet boundary condition with 
\be
A_\m=0~.
\ee
Neglecting the radial component this is then $A_i=0$.

\end{document}